\documentclass[aps,prb,twocolumn,superscriptaddress,floatfix,citeautoscript,longbibliography]{revtex4-2}

\usepackage{verbatim}
\usepackage{latexsym,amssymb,float}
\usepackage{setspace}
\usepackage{graphicx}
\usepackage{dcolumn}
\usepackage{epsfig}
\usepackage{epstopdf}
\usepackage{amsmath}
\usepackage{bm}
\usepackage{tikz}
\usepackage{caption}
\usepackage{subcaption}
\normalsize 
\usepackage{booktabs}
\usepackage{braket}
\usepackage{times}
\usepackage{bbold}
\usepackage[colorlinks=true]{hyperref}

\newcommand{\be}{\begin{equation}}
\newcommand{\ee}{\end{equation}}
\renewcommand{\vec}[1]{\bm{#1}}

\newcommand{\bea}{\begin{eqnarray}}
\newcommand{\eea}{\end{eqnarray}}

\newcommand{\ra}{\rangle}

\newcommand{\ri}{\mbox{i}}

\renewcommand{\vec}[1]{{\bm #1}}

\begin{document}

\title{A mesoscopic device for a realization of the Topological Kondo effect}
\author{Saheli Sarkar}
\author{Alexei M. Tsvelik}

\affiliation{
    Division of Condensed Matter Physics and Materials Science, Brookhaven National Laboratory, Upton, NY 11973-5000, USA}

\begin{abstract}
The search for anyons is a field of immense interest owing to its potential application in the field of quantum information. Quantum critical Kondo impurities constitute one possible  platform for their realization and  Topological Kondo effect (TKE) by virtue of remaining  critical in the presence of perturbations, seems to be especially promising in this regard. In this paper we discuss practical steps for a realization of TKE with a relatively high Kondo temperature $T_K$. Its central feature is the so-called Majorana-Cooper box (MCB) and we argue that a particular type of  iron-based topological superconductor is especially suitable for realization of TKE. Once MCB is available one needs to  connect it  to external metallic leads to produce TKE. A relatively high value of the Kondo temperature $T_K$ is then aided by a large superconducting gap of the iron-based superconductor. We give estimates for $T_K$, for the cases of both isotropic and anisotropic exchange couplings of MCB with the leads.
\end{abstract}


\maketitle

\section{Introduction}
 
  Anyons - fractionalized particles with exotic statistics, are considered an essential part of quantum information systems \cite{Sankar2008}. There are different suggestions of their realization, one of them being the Kondo effect. 
  Existence of quantum critical Kondo effect with a non-Fermi liquid (NFL) behavior, first described and studied theoretically \cite{Nozieres1980, TsvW1984, AfflLud} has been well established experimentally \cite{potok2007,keller2015,Iftikhar2015,Iftikhar2018,karski2023}. It is  well understood that the corresponding critical ground state contains non-Abelian anyons \cite{Sela2020,Komijani2020,Ge2022}.  However, the experiments implement  the so-called multichannel Kondo effect where the criticality is sensitive to  asymmetry between the scattering channels. Hence a realization of the quantum critical point requires  a fine tuning of the coupling parameters which  limits its potential applications. 
 
  B\'{e}ri and Cooper \textit{et.al.} \cite{BeriCooper2012} proposed a new type of non-Fermi liquid Kondo effect where the criticality is stable against most common perturbations.  The Topological Kondo effect (TKE)  is predicted to arise as a result of coupling between bulk conduction electrons and  Majorana Zero modes [MZM,  also called Majorana Bound States (MBSs) ] located inside of the so-called Majorana-Cooper box (MCB). They also provided a sketch  of the MCB setup for TKE  
 \cite{BeriCooper2012} in the form of a device consisting of a mesoscopic superconducting island  with proximitized to it  semiconductor nanowires containing the MBSs. The Coulomb blockade in the island transforms entangled MBSs into an effective quantum spin, which interacts with the conduction electrons of the leads via superexchange interaction  facilitating the TKE. In contrast to the conventional quantum critical Kondo effect, the TKE is 
  robust against various perturbations \cite{TopKondo,TopKondoWe}.

 Apart from being an anyon platform,  the  TKE with its characteristic features may  resolve persistent controversies about the existence of MBS. The theory \cite{BeriCooper2012,TopKondo} suggests that in the device with $M$ leads, the linear conductance ($\sigma_{ij}$) between different leads $i$ and $j $ saturates at small temperatures at the  universal value $\frac{2e^2}{M h}$, following a nontrivial power law temperature dependence, different from the $T^2$ Fermi liquid one. This universal behavior of the conductance constitutes a manifestation of the anyonic character of the TKE ground state, which therefore can serve as a smoking gun for the presence of MBS. 

 The aim of this paper is to provide detailed suggestions for realization of MCBs for TKE focusing on  iron-based superconductors (FeSC) based systems, with a final goal to use it for quantum information applications.
We provide different scenarios for realizing MCBs using FeSC and parameters characterizing the device for realizing TKE, in the absence and presence of anisotropy of the exchange couplings between MCB and the external metallic leads. We also depict an arrangement for a TKE-based chiral Kondo lattice which contains multiple non-Abelian anyons.

  As we have mentioned above, the suggestions of realization of TKE contained in earlier theoretical studies described  heterostructure devices containing nanowires with strong  spin-orbit interaction in the presence of magnetic field proximitized  to a mesoscopic size superconducting island  \cite{BeriCooper2012,TopKondoWe,Roman2020,Dong2021}. We suggest another route to TKE:  to use  intrinsic topological superconductor (TSC) \cite{Beenakker2013,Sato2017topo,Zhang2016} based devices, which may naturally host Majorana fermions. This may drastically simplify the fabrication process providing a material realization for the TKE. Promising candidates include   iron-based superconductors 
 Fe(Te$_x$,Se$_{1-x})$ ($T_c = 14.5$ K)  \cite{Wang2018, Zhang2018, Machida2018}, (Li$_{0.84}$Fe$_{0.16}$)OHFeSe ($T_c =42$ K) \cite{Liu2018} and  CaKFe$_2$As$_4$ ($T_c =35$ K) \cite{Liu2020}. All of them 
 exhibit signatures of intrinsic topological superconductivity and have relatively high superconducting critical temperatures T$_c$. 
 Thus, it is conceivable that MCB  based on such materials  can provide  a viable path towards observing the TKE. Observation of TKE in turn can serve as a smoking gun signature of MBS, whose detection in various  systems has been so far largely focused on observing a zero-bias conductance peak (ZBCP)\cite{Yin2015,Zhu2020} in local probe techniques. Additionally, finding signatures of non-local character \cite{Alicea2012} of MBS through TKE will be important for quantum computation applications.

 \section{Different scenario for realization of Majorana-Cooper boxes}
 
    There are multiple experiments indicating a possible existence of MBSs in various iron-based superconductors. Non of them are decisive, but as we suggest, in the context of TKE the corresponding systems may provide smoking gun evidence for MBS. Below we consider different arrangements weighting their merits as possible candidates for realization of TKE. In all cases the corresponding MCBs are conceived as mesoscopic devices consisting of superconductor with charging energy $E_C$ containing MBSs.

 
 \textit{Vortices:} One way of coexistence of MBSs and superconductivity is through vortices. There are multiple observations of ZBCP trapped in vortex cores of superconductors Fe(Te$_x$,Se$_{1-x})$ \cite{Wang2018, Zhang2018, Machida2018}, (Li$_{0.84}$Fe$_{0.16}$)OHFeSe \cite{Liu2018} and  CaKFe$_2$As$_4$  \cite{Liu2020}. These experiments are interpreted as evidence for MBS. However, there are several difficulties in clearly distinguishing the MBSs, for e.g. due to the presence of topologically trivial Caroli-de Gennes- Matricon \cite{CAROLI1964} states at the vortex cores. The Caroli-de Gennes- Matricon states are observed  to produce a broad peak centering at the zero energy, as the energy separation between the states is of the order of $\Delta^2/E_F$, where $E_F$ is the Fermi energy. Thus as the energy separation can be significantly smaller than the instrumental energy resolution, they appear as a broad peak at the zero energy, being different in origin from that of the MBS. Moreover, scanning-tunneling microscopy (STM) experiments \cite{Machida2018,kong2019half} in FeSe$_{0.45}$Te$_{0.55}$ have reported the presence of the ZBCP only in a fraction of the vortex cores present in the system, depending on the magnetic field, rendering the MBSs questionable.

 
 \textit{Magnetic point defects:} Quantum anomalous vortices \cite{Jiang2019} may nucleate at magnetic impurities at zero magnetic field. In the presence of a surface topological superconductivity they can support MBSs at the vortex centers, manifesting themselves as ZBCP in tunneling experiments. These types of sharp zero energy peaks located at interstitial iron impurities (IFI) have been observed by STM in the superconducting state of Fe$_{1+x}$(Te,Se) \cite{Yin2015} with superconducting gap of about 2 meV, below superconducting temperature $T_c = 14.5$ K. The zero bias peak intensity decays exponentially from the center of the peak with correlation length $\xi \approx 3.5$ \AA, which is much smaller than the superconducting correlation length $\sim 25$ \AA. The STM have found IFIs on the exposed $(Te,Se)$ surface  located right at the mid position between  four neighboring (Te,Se) atoms. They manifest themselves as sharp peaks at zero bias; the peaks remain robust in applied magnetic field up to 8 T. Similar ZBCP have also been found later in LiFeAs \cite{Zahid2020} and in IFI on monolayers of FeSe and FeSe$_{0.5}$Te$_{0.5}$ on /SrTiO$_3$(STO) \cite{liu2020zero}. One potential problem is that  the peak has an intrinsic width in all of the systems [for e.g., in  Fe$_{1+x}$(Te,Se) \cite{Yin2015} the width $\sim 0.6$ meV at $T= 1.5$ K] whose origin is unclear. Perhaps, this widening can be taken into account phenomenologically as a result of a coupling of the MBS to the bath. In any case, it will likely to have a damaging effect on TKE. Additionally, in a STM experiment \cite{Fan2021} in FeSe$_{0.45}$Te$_{0.55}$ with IFI, the ZBCP was found only near some  iron adatoms, the rest of them having Yu-Shiva- Rusinov (YSR) bound states at finite energies.


\textit{Line defects:} The STM study \cite{Chen2020} reports a discovery of zero-energy bound states simultaneously appearing at both ends of a one-dimensional
atomic line defect in monolayer iron-based high-temperature superconductor FeTe$_{0.5}$Se$_{0.5}$ films grown on SrTiO$_3$(001) substrates 
, which has $T_c \approx 62$ K. These line defects naturally emerge during the growth process and correspond to lines of missing Te/Se atoms. 

 \begin{figure}[t]
\includegraphics[width=0.45 \textwidth]{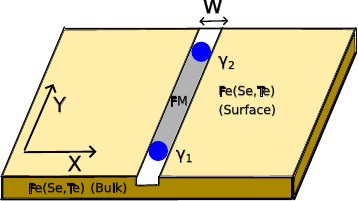}
\caption{A schematic representation of a mesoscopic setup for the formation of a paired Majorana bound states: The device contains a slab of Fe(Se,Te) superconductor and a ferromagnetic film (FM) of width W deposited on the surface of the superconductor, with W being much smaller than the size of the superconductor. This effectively gives rise to a superconductor-ferromagnet-superconductor (S-F-S) junction. \label{Fig:SFS}}
\end{figure}

Another recent STM \cite{Madhavan2020} study has found evidence of Majorana fermions in a certain type of crystalline domain walls associated to the half-unit cell shift of the Se atom in superconducting FeSe$_{0.45}$Te$_{0.55}$ [Fe(Se,Te)]. It was established in \cite{Johnson}, this material also develops surface ferromagnetism. The first principle calculations \cite{Song2022} have suggested that these types of crystalline domain walls can develop an in-plane ferromagnetism along the domain wall orientation and may support Majorana modes. Moreover, it has been also suggested, that if the magnetization can be tuned, MBS can be trapped by defects such as ferromagnetic domain wall \cite{Song2022}. 
 
 Motivated by these, we consider a model device consisting of a mesoscopic Fe(Se,Te) and a ferromagnetic film, which can be used to study TKE with a higher Topological Kondo temperature scale, thus providing advantage of previously proposed setup based on complex heterostructures.
 
Below we describe in detail the setup of a device giving rise to a pair of MBS's. It is composed of a slab of Fe(Se,Te) material and a thin ferromagnetic film (FM) deposited on the surface of the Fe(Se,Te) or any similar material as shown in the Fig.\ref{Fig:SFS}. Below the T$_c$, one part of the device becomes superconducting, in the thin film area, however, the ferromagnetism persists. Hence, the system can be simply modeled as a superconductor-ferromagnet-superconductor (S-F-S) junction [see Fig.\ref{Fig:SFS}]. In the  Fig. \ref{Fig:SFS}, the x-axis points to the normal to the junction between the ferromagnet and the superconductor, where as the y-axis points parallel to the orientation of the junction. The semi-infinite superconducting regions occupy intervals  $x< -W/2$ and $x > W/2$, while the FM region occupies interval $- W/2 <x<W/2$. We consider that the system size along y-direction as infinite, the width W remains finite. The Fe(Se,Te) is known to have non-trivial topological spin-helical Dirac surface states \cite{Zhang2018} as well as the Rashba type spin-orbit coupling \cite{Rafael2019,Mascot2022}. The bulk of Fe(Se,Te) possesses a node-less and almost isotropic s-wave type superconducting gap \cite{Sarkar2017}. The FM produces an effective Zeeman field $h_y$ parallel to the junction. 
  
Consequently, the surface S-F-S junction of width W is described by the following Dirac-Bogolyubov-de-Gennes (BdG) Hamiltonian \cite{Song2022} in the Nambu basis $\Psi = (\psi^{\dagger}_{k, \uparrow},\psi^{\dagger}_{k, \downarrow}, \psi_{-k,\uparrow},\psi_{-k, \downarrow})$ :
 
\begin{align}\label{Ham:BdG}
H_{BdG} &=
\begin{pmatrix}
 H_{0}(k) & i \sigma_{y}\Delta(r)\\ 
  -i \sigma_y \Delta^*(r) & -H_{0}^{*}(-k).
\end{pmatrix}
\end{align}
In Eqn.\eqref{Ham:BdG}, $H_{0}(k)$ is a $2 \times 2$ matrix, $H_{0}(k) = H_{SO} - \mu +H_{Z}(x)$. The $H_{0}(k)$ contains the Rashba spin-orbit coupling term  $H_{SO}=\alpha (\vec{\sigma} \times \vec{k}). \hat{z}$, associated to the Dirac surface states of the Fe(Se,Te), a chemical potential $\mu$ and a Zeeman energy term  due to the FM given by $H_{Z}(x)=\sigma_y h_y \Theta[(W/2) - |x|]$. Here, $\alpha = v_{F} \hbar$, is the coupling strength, and $v_F$ is the Fermi velocity. $\sigma$ are the Pauli matrices in the spin- space. With $\hbar = 1$,  the Rashba term $H_{SO}$ becomes $v_F (\sigma_x k_y +i \sigma_y \partial_{x} )$, where $k_y$ is momentum along the y-direction that is conserved in the system. The spatial variation of the superconducting gap function on the surface satisfies 
$\Delta(r) = \Delta_{0} e^{isgn(x)(\theta/2)}\Theta [|x| - W/2]$, where $\theta$ is the phase difference between the left and the right superconducting regime.

It has been shown in \cite{Song2022}, that the S-F-S junction described by the Hamiltonian Eqn.\eqref{Ham:BdG} can support one dimensional counter propagating Majorana modes for a constant phase difference ($\Delta \phi=\pi$) between the two superconducting regimes on the left and right arising due to the in-plane magnetic moment in the FM. Moreover, the localized MBSs are formed \cite{Kane2008}from the coupling between these dispersive Majorana modes, if the phase difference $\Delta \phi (y)$ varies continuously along the orientation $y$ of the junction such that, $\Delta \phi (y)$ becomes an odd integer multiple of $\pi$. This follows from the analysis of  the low energy effective Hamiltonian derived from the BdG Hamiltonian Eqn.\eqref{Ham:BdG}, by projecting it onto the basis of the one-dimensional Majorana modes \cite{Song2022}:
\begin{eqnarray}
&& H_{eff} = v_m k_y \tau_{y} - \Delta_{0} \cos \left[\frac{\Delta \phi (y)}{2}\right]\tau_{z},\label{Ham:Loweff}\\
&& \Delta \phi (y) = \frac{2 h_y W}{v_F},\label{Ham:phase}\\
&&  v_m  = v_{F} \left[\cos k_F W + \frac{\Delta_0}{\mu} \sin k_F W\right]\frac{\Delta^2_{0}}{(\mu^2 + \Delta^2_{0})}\label{vm}
\end{eqnarray}

, where $\tau$ are the Pauli matrices in the particle-hole space. Note that the Eqn.\eqref{Ham:Loweff} has the form of a Su-Schrieffer-Heeger model. It is also well-known from the Jackiw-Rabbi problem, that a zero energy localized MBS forms when the mass term $ \Delta_{0} \cos (\frac{\Delta \phi (y)}{2})$ in Eqn.\eqref{Ham:phase} smoothly changes sign or $\Delta \phi (y)$ becomes an odd integer multiple of $\pi$. The Bogoliubov quasiparticle operator associated with this state is a Majorana fermion, which satisfies $\gamma_{0} = \gamma^{\dagger}_{0}$. 

Therefore in this setup, to tune the phase difference $\Delta \phi (y)$ for obtaining a pair of MBSs, either the width $W$ of the FM can be changed along the $y$ direction or a soft magnetic material for the FM can be used where one can vary the magnetization strength $h_y$.

\begin{figure}[t]
\includegraphics[width=0.52 \textwidth]{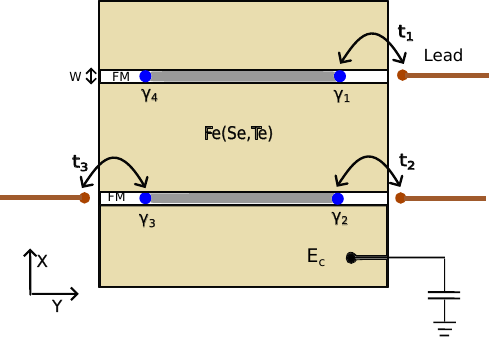}
\caption{A schematic representation of a  mesoscopic setup for Topological Kondo effect with $N=4$ MBSs and $M=3$ metallic leads: The device contains a Coulomb blockaded superconducting island made of Fe(Se,Te) superconductor and FM. The island is connected to the ground through a capacitor and hence possesses a charging energy $E_c$. 3 MBSs are tunnel coupled to the metallic leads with coupling constants $t_{j}$'s (j= 1,2,3). }
\label{Majorana-Coulomb}
\end{figure}

\section{A $\textbf{FeSC}$ -based device for observation of Topological Kondo effect}
The MCB device considered in our work for the realization of the Topological Kondo effect (TKE) comprises of a mesoscopic Fe(Se,Te) (or a similar material) and two ferromagnetic films deposited on it and resulting in a pair of MBS in each of the ferromagnetic films as shown in Fig. \ref{Majorana-Coulomb}. Realization of the TKE suggested in the Ref.\cite{BeriCooper2012} is based on the MCB which was described, for example, in \cite{Liu2019}, containing $N \geq 4$ MZMs. Following the suggestions, we consider a floating mesoscopic MCB island, grounded through a capacitor. To get the critical ground state, one needs to connect MZMs to just three external metallic leads (this is a minimal number) via tunneling contacts as schematically depicted in Fig. \ref{Majorana-Coulomb}.  The charging energy $E_C= \frac{e^2}{2C}$, with $C$ being the geometric capacitance of the box, contributes to the Hamiltonian of the MCB island as 
 
 \be\label{eq:charging}
 H_c = E_c (N -Q_0/e)^2.
 \ee
In Eqn.\eqref{eq:charging}, $N$ is the number of electrons in the island 
 and $Q_0$ is the background charge determined by the voltage across the capacitor, connected to the MCB.

In general, if there are a total N number of MBS in the MCB, there will be $N/2$ number of zero energy fermionic modes. As the total charge of the island is fixed by a large charging energy $E_c$, to even or odd number, thus gives a parity constraint to the MBS operators. This leads to a ground state of the system with $2^{(N/2)-1}$ fold degeneracy. In this work, we consider the arrangement where MCB island contains four Majorana zero modes $N=4$, hence the ground state is two-fold degenerate. The two degenerate quantum ground states,  $|\downarrow\ra$ ($|\uparrow\ra$),  have $N_0$ ($N_0-2$) particles in the condensate and empty (filled) pairs of Majorana modes, thus encoding a qubit~\cite{Fu2010,BeriCooper2012}. It is this topological degeneracy which will lead to the effective spin degeneracy for the effective Kondo model. Now, we assume, no overlap between the MBS, hence the degeneracy remains exact, otherwise a ``Zeeman coupling" \cite{TopKondoWe} term can arise.  

The TKE occurs when the MCB is coupled to conduction electrons through external metallic leads as shown in Fig. \ref{Majorana-Coulomb}. We work at the energy scales much smaller than the superconducting gap, as well as below the energies of any other subgap excitations of non-Majorana character. At such temperatures the subgap states are not populated and can be ignored. Then, the Hamiltonian of the system is given by $H_{TK} = H_{cond} + H_{c} + H_{tun}$, where $H_{cond}$ is the Hamiltonian of the conduction electrons of the leads. The tunneling Hamiltonian $H_{tun}$ describes the low-energy coupling between the leads and the MCB island. $H_{tun}$ is given by
 \bea
 H_{tun} = \exp(\ri\phi/2)\sum_{j=1}^{3}t_{jj}\gamma_j\psi_j + \mbox{H.c.}, \label{tunn}
 \eea 
  where $\phi$ is the phase of the superconducting order parameter, $t_{jj}$ is the tunneling amplitude, $\psi_j$ is the electron annihilation operator and $\gamma_j$ is the Majorana operator at lead j. Such tunneling process explicitly excludes the possibility of exciting quasiparticles \cite{Plugge2}. As was demonstrated in \cite{BeriCooper2012,Galpin2014}, if the charging energy of the box is much greater than the characteristic value of the tunneling matrix elements, i.e. if $E_C >> t_{j}$, one can integrate out the phase fluctuations of the condensate which results in the exchange interaction between such MCB and the electrons of the leads (it is supposed that they are spin polarized): 
 \bea
H_{ex} =  J_K^{ij}(\psi^+_i\psi_j - \psi^+_j\psi_i)\gamma_i\gamma_j, ~~ J_K^{ij} \sim t_it_j/E_C, \label{Kondo}
 \eea
  where indices $i,j$ correspond to the leads. For a single MCB this interaction gives rise to the TKE, where the leads serve as the bulk.  The spin operator $S^i$ realized by the Majorana operators is 
  \bea
  S^i = \frac{\rm i}{2}\epsilon_{ijk}\gamma_j\gamma_k, ~~ \{\gamma_k,\gamma_j\} = \delta_{jk}. \label{rep1}
  \eea
The model comprised  of the lead Hamiltonian and the effective exchange Hamiltonian [Eqn.\eqref{Kondo}]  defines an antiferromagnetic Kondo problem.

In the limit of isotropic couplings, where all of the $J^{ij}_K$s are same, implying $J^{12}_K = J^{23}_K=J^{13}_K$, one arrives at the RG flow equations  of the isotropic Kondo problem. In this limit, the Topological Kondo temperature $T_K$ at which the perturbation theory breaks down due to the RG flow  towards strong coupling is given by \cite{BeriCooper2012,TopKondo,TopKondoWe},
\bea \label{Eq:Tk}
T_K \sim E_c \exp(-1/\rho \bar J_K), 
\eea
where $\rho$ is the density of states of the lead electrons at the Fermi energy and $J^{12}_K = J^{23}_K=J^{13}_K =\bar J_K$ is the average bare value of the exchange coupling. Consequently, $T_K$ sets an energy scale between the trivial regime and the Topological Kondo regime, below which a robust non-Fermi liquid (NFL) behavior can be observed. However, a strong exponential factor in Eqn.\eqref{Eq:Tk} can drive the temperature for the TKE to zero.
In deriving $T_K$ [Eqn.\eqref{Eq:Tk}], a possibility of hybridization between the MBS has been ignored. If the hybridization of the form $H_{hyb} = i \sum_{j\neq k} h_{jk} \gamma_j\gamma_k$ is present, the TKE will be then observed in a temperature window $T_{hyb} < T < T_K$, where $T_{hyb} = T_K (\bar h /T_K)^{M/2}$ \cite{TopKondoWe}, with $\bar h = max |h_{jk}|$. 

Anisotropies \cite{TopKondoWe} in the exchange couplings are marginal  perturbations so the NFL fixed point of TKE remains  robust  \cite{TopKondo,zazunov2014transport}. 
In the following, we consider the anisotropic case when $J^{12}_K = J^{13}_K \neq J^{23}_K$. In the new notations  $J^{12}_K = J^{13}_K \equiv J_{\perp}$ and $ J^{23}_K \equiv J_{\parallel}$. The one-loop perturbative RG equations in this case are, 

\begin{align}\label{rgeq1}
\frac{dJ_\parallel}{d \log \Lambda} &= -\rho J_{\perp}^2 \\
\frac{dJ_{\perp}}{d \log \Lambda} &= -\rho J_{\parallel}J_{\perp}.    
\end{align}

From the above RG equations Eqn.\eqref{rgeq1} one finds that,
\begin{align}\label{eqnscale}
\nonumber
\frac{dJ_{\parallel}}{dJ_{\perp}} & = \frac{J_{\perp}}{J_{\parallel}}\\
\Rightarrow J^2_\parallel - J^2_{\perp} &= C
\end{align}

Eqn.\eqref{eqnscale} represents a scaling constraint to be obeyed, where $C$ is a real constant. Next, plugging in Eqn.\eqref{eqnscale} in Eqn.\eqref{rgeq1}, we obtain
\begin{align}\label{rgeqn2}
\frac{dJ_\parallel}{d \log\Lambda} & = -\rho (J^2_\parallel - C).
\end{align}
To obtain the Topological Kondo energy scale, we integrate the above RG equation [Eqn.\eqref{rgeqn2}]. Following the approach in \cite{BeriCooper2012}, the energy cutoff or the effective bandwidth $\Lambda$ varies between $[T_K, D]$, where $D \sim E_c$, where $E_c$ is the charging energy of the island. There are two regimes $C>0$ and $C<0$. $T_K$ will give the Topological Kondo temperature, at which the coupling constants flow to strong-coupling which for the estimate purposes we can treat as infinity. 

We assume that $\rho\sqrt |C| <<1$. Then for $C>0$ we obtain
\bea
T_K \sim E_c\Big(\frac{-\sqrt C + J_{\parallel}(0)}{\sqrt C + J_{\parallel}(0)}\Big)^{1/2\rho\sqrt C},
\eea
Here, we see that to have a greater $T_K$ it is advantageous for us to have strong anisotropy $J_{\parallel} >> J_{\perp}$ which corresponds to the situation when one of the tunneling integrals $t_i$ is much smaller than two others. 

For $C<0$ we have 
\bea
T_K \sim E_c\exp\Big[\frac{-1}{\rho \sqrt{|C|}}(\pi/2 - \tan^{-1}(J_{\parallel}(0)/\sqrt{|C|})\Big]
\eea
, where $J_{\parallel}$(0) is the bare coupling.

 One problem is that in all cases apart from tunneling into MBS (\ref{tunn}) there may be a direct tunneling of bulk electrons into the superconductor. There are, however, reasons why this process is irrelevant. 
Indeed, the corresponding  Hamiltonian is 
\bea
H_{tun,2} = g\Big[A_{lj}\exp(\ri\phi)\psi_l\psi_j   + H.c.\Big], ~~ A_{lj} = - A_{jl}.
\eea
When we integrate over $\phi$, however, the result is a local four fermion interaction 

\bea
V \sim \frac{g^2}{E_c}A_{lj}A_{l'j'}\psi_l\psi_j\psi^*_{l'}\psi^*_{j'},
\eea
which is highly irrelevant.
  
Below we give estimates of the parameters pertinent to the realization of TKE for the device setup depicted in Fig.\ref{Majorana-Coulomb}.

 We start by summarizing the criteria to be satisfied are: (i) four well-localized MBS, (ii) the MBS must be  well-separated, such that the hybridization between them is insignificant, but fit inside the Coulomb box,  (iii) a charging energy $E_c$ of the order of the superconducting gap $\Delta$. These conditions can be expressed in the sequence of inequalities:
 \begin{equation}
     \Delta_0 >> E_c >> t_i, ~~ C \sim L >> \lambda. 
 \end{equation}
The decay length of the MBS is given by $\lambda = \frac{v_m}{\Delta_{0}}$ \cite{Song2022}, where $v_m$ is given by Eq.(\ref{vm}).  To make it to be much smaller than the size (L) of the Coulomb box we need to reduce $v_m$. For FeSe$_{0.45}$Te$_{0.55}$, with typical values of $v_F \sim 216$ meV \AA \cite{Wang2018}, $\Delta_0 \sim 1.8$ meV, $\mu \sim 5 \Delta_0$~($\mu >> \Delta_0$), the smallest value of $v_m \approx v_F(\Delta_0/\mu)^3$, $\lambda$ of order of 1 \AA~is achieved when $k_FW = \pi(1+2n)/2$. It is important to note that the decay length $\lambda$ is well below the superconducting coherence length of the Fe(Se,Te)\cite{audouard2015quantum,singh2015evidence}, which is required for the stability of the MBS.

With a superconducting island of area $A \sim 60$ nm $\times 60$ nm and thickness d $\sim 1$ nm, and assuming permittivity of an insulating material being $\epsilon \sim 2 \times 55  e^2 ev^{-1} (\mu m)^{-1} $, the charging energy $E_c = \frac{e^2}{2 C}$, with $C = \frac{\epsilon A}{d}$  will be of the order of 1.38~meV $\equiv 16  K \sim \Delta_0$. Hence we notice that, the device geometry can achieve a large upper-bound for the topological Kondo temperature scale, while remaining in the superconducting state. We also see that the effective Topological Kondo temperature is determined not just by $E_c$, but also by the tunneling integrals $t_i$ which enter into the expressions for the exchange integrals. Thus, even a suppression due to the exponential factor, can result in a relatively larger $T_K$, as long as $E_c$ is large. The suppression will depend on the density of states of the metallic lead-electrons and the nature of the contacts used.
Therefore, we see that a mesoscopic MCB island with well-separated and localized MBS can be achieved with a large $T_K$ [see for e.g. Eqn.\eqref{Eq:Tk}]. As the $T_K$ achieved in the Fe(Se,Te) based device can be significantly higher than the heterostructures systems based on semiconductor nanowire and conventional superconductors \cite{BeriCooper2012}, this device will significantly improve the experimental feasibility of the TKE.

 As we have mentioned in the introduction, that one possible way to detect the TKE is to measure the conductance between different leads, as the TKE leads to unusual temperature dependencies   for the two-terminal linear conductance $\sigma_{ij}$ ($i \neq j$)\cite{TopKondo,TopKondoWe}. In the temperature-regime larger than  $T_K$, the $\sigma_{ij}$ grows as $\frac{1}{ln^2(T/T_K)}$ with lowering the temperature. For temperature-regime smaller  than $T_K$,  $\sigma_{ij}$ approaches the universal value $\frac{2e^2}{3h}$ with $(T/T_K)^{2/3}$ temperature dependence. Both the universal value and the temperature dependence of the conductance are unique characteristics of  the NFL  ground state  
providing an alternative pathway for detecting the Majorana fermions in the Fe(Se,Te) based device. Moreover, the signature of the TKE can also be found in the tunneling measurements for example using STM. Ref. \cite{ErikssonPRB2014} describes the situation when the leads are one-dimensional wires with the Luttinger parameter $K$. In the presence of the TKE, the local density of states (LDOS) $\rho(\omega)$ of the lead-electrons close to the MCB island, as a function of the energy has a dependence $\rho(\omega) \sim \omega^{\frac{1}{MK}[1+(M-1)K^2]-1}$\cite{ErikssonPRB2014}, where $K$ is related to the Coulomb repulsion between electrons in the lead. Hence, depending on the strength of the interaction among the electrons in the metallic lead, the LDOS will show unique features, which can provide an alternative experimental route to find signature of the TKE.

\begin{figure}[!htb]
\centerline{\includegraphics[ angle = 0,
width=0.5\columnwidth]{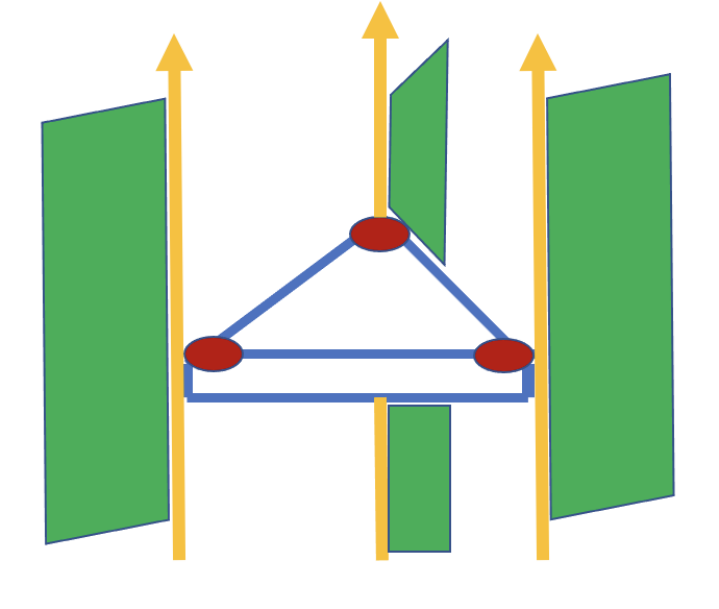}}
\vspace{-.50cm}
\caption{A schematic depiction of the arrangement for the Chiral Kondo effect. The  triangle is the superconducting (Cooper pair) box, the red dots are Majorana zero modes  coupled by tunneling to chiral edges of topological insulators.   
}
 \label{KondoChiral}
\end{figure}
\textit{Arrays of TKE:} We conclude our paper with a brief description of a device which can incorporate multiple anyons. This is chiral Kondo lattice first described in \cite{Sela2020,Sela2022,Lotem2022}. Here the itinerant electrons are chiral, they move in one direction and hence such  lattice does not have Ruderman–Kittel–Kasuya–Yosida (RKKY) interaction. In fact, it is equivalent to the bunch of impurities and as a consequence  the ground state remains critical as for the single impurity case.

Chiral Kondo lattice for Topological Kondo effect can serve as a platform for anyon-based quantum computation \cite{Sela2020,Sela2022,Lotem2022}. An elementary block of such Kondo lattice is depicted on Fig.(\ref{KondoChiral}). The suitable material for chiral edges containing polarized electrons is MnBi$_2$Te$_4$ \cite{MnBiTe}, which is distinguished by a relatively large bulk gap, allowing one to operate at temperatures of order of several degrees. 

\section{Summary}
In this work, we have provided a new potential route to achieve the Topological Kondo effect (TKE) through various iron-based superconductors such as FeSe$_{0.45}$Te$_{0.55}$ [Fe(Se,Te)] based Majorana-Cooper boxes (MCB). The proposed device set up has two marked differences from the earlier proposals set ups which include heterostructures containing semiconductor nanowires proximitized to conventional superconductors. First, there is a significant enhancement of the Topological Kondo temperature scale and the second, there is a simplification of the device structure. 

\section{Acknowledgements}
This work was supported by U.S. Department of Energy (DOE) the Office of Basic Energy Sciences, Materials Sciences and Engineering Division under Contract No. DE-SC0012704.

\bibliography{TKE}

\providecommand{\noopsort}[1]{}\providecommand{\singleletter}[1]{#1}%
\begin{thebibliography}{53}%
\makeatletter
\providecommand \@ifxundefined [1]{%
 \@ifx{#1\undefined}
}%
\providecommand \@ifnum [1]{%
 \ifnum #1\expandafter \@firstoftwo
 \else \expandafter \@secondoftwo
 \fi
}%
\providecommand \@ifx [1]{%
 \ifx #1\expandafter \@firstoftwo
 \else \expandafter \@secondoftwo
 \fi
}%
\providecommand \natexlab [1]{#1}%
\providecommand \enquote  [1]{``#1''}%
\providecommand \bibnamefont  [1]{#1}%
\providecommand \bibfnamefont [1]{#1}%
\providecommand \citenamefont [1]{#1}%
\providecommand \href@noop [0]{\@secondoftwo}%
\providecommand \href [0]{\begingroup \@sanitize@url \@href}%
\providecommand \@href[1]{\@@startlink{#1}\@@href}%
\providecommand \@@href[1]{\endgroup#1\@@endlink}%
\providecommand \@sanitize@url [0]{\catcode `\\12\catcode `\$12\catcode
  `\&12\catcode `\#12\catcode `\^12\catcode `\_12\catcode `\%12\relax}%
\providecommand \@@startlink[1]{}%
\providecommand \@@endlink[0]{}%
\providecommand \url  [0]{\begingroup\@sanitize@url \@url }%
\providecommand \@url [1]{\endgroup\@href {#1}{\urlprefix }}%
\providecommand \urlprefix  [0]{URL }%
\providecommand \Eprint [0]{\href }%
\providecommand \doibase [0]{https://doi.org/}%
\providecommand \selectlanguage [0]{\@gobble}%
\providecommand \bibinfo  [0]{\@secondoftwo}%
\providecommand \bibfield  [0]{\@secondoftwo}%
\providecommand \translation [1]{[#1]}%
\providecommand \BibitemOpen [0]{}%
\providecommand \bibitemStop [0]{}%
\providecommand \bibitemNoStop [0]{.\EOS\space}%
\providecommand \EOS [0]{\spacefactor3000\relax}%
\providecommand \BibitemShut  [1]{\csname bibitem#1\endcsname}%
\let\auto@bib@innerbib\@empty
\bibitem [{\citenamefont {Nayak}\ \emph {et~al.}(2008)\citenamefont {Nayak},
  \citenamefont {Simon}, \citenamefont {Stern}, \citenamefont {Freedman},\ and\
  \citenamefont {Das~Sarma}}]{Sankar2008}%
  \BibitemOpen
  \bibfield  {author} {\bibinfo {author} {\bibfnamefont {C.}~\bibnamefont
  {Nayak}}, \bibinfo {author} {\bibfnamefont {S.~H.}\ \bibnamefont {Simon}},
  \bibinfo {author} {\bibfnamefont {A.}~\bibnamefont {Stern}}, \bibinfo
  {author} {\bibfnamefont {M.}~\bibnamefont {Freedman}},\ and\ \bibinfo
  {author} {\bibfnamefont {S.}~\bibnamefont {Das~Sarma}},\ }\bibfield  {title}
  {\bibinfo {title} {Non-abelian anyons and topological quantum computation},\
  }\href {https://doi.org/10.1103/RevModPhys.80.1083} {\bibfield  {journal}
  {\bibinfo  {journal} {Rev. Mod. Phys.}\ }\textbf {\bibinfo {volume} {80}},\
  \bibinfo {pages} {1083} (\bibinfo {year} {2008})}\BibitemShut {NoStop}%
\bibitem [{\citenamefont {Nozieres}\ and\ \citenamefont
  {Blandin}(1980)}]{Nozieres1980}%
  \BibitemOpen
  \bibfield  {author} {\bibinfo {author} {\bibfnamefont {P.}~\bibnamefont
  {Nozieres}}\ and\ \bibinfo {author} {\bibfnamefont {A.}~\bibnamefont
  {Blandin}},\ }\bibfield  {title} {\bibinfo {title} {Kondo effect in real
  metals},\ }\href {https://doi.org/10.1051/jphys:01980004103019300} {\bibfield
   {journal} {\bibinfo  {journal} {Journal de Physique}\ }\textbf {\bibinfo
  {volume} {41}},\ \bibinfo {pages} {193} (\bibinfo {year} {1980})}\BibitemShut
  {NoStop}%
\bibitem [{\citenamefont {Tsvelick}\ and\ \citenamefont
  {Wiegmann}(1984)}]{TsvW1984}%
  \BibitemOpen
  \bibfield  {author} {\bibinfo {author} {\bibfnamefont {A.}~\bibnamefont
  {Tsvelick}}\ and\ \bibinfo {author} {\bibfnamefont {P.}~\bibnamefont
  {Wiegmann}},\ }\bibfield  {title} {\bibinfo {title} {Solution of the
  n-channel kondo problem (scaling and integrability)},\ }\href
  {https://link.springer.com/article/10.1007/BF01319184} {\bibfield  {journal}
  {\bibinfo  {journal} {Zeitschrift f{\"u}r Physik B Condensed Matter}\
  }\textbf {\bibinfo {volume} {54}},\ \bibinfo {pages} {201} (\bibinfo {year}
  {1984})}\BibitemShut {NoStop}%
\bibitem [{\citenamefont {Affleck}\ and\ \citenamefont
  {Ludwig}(1991)}]{AfflLud}%
  \BibitemOpen
  \bibfield  {author} {\bibinfo {author} {\bibfnamefont {I.}~\bibnamefont
  {Affleck}}\ and\ \bibinfo {author} {\bibfnamefont {A.~W.}\ \bibnamefont
  {Ludwig}},\ }\bibfield  {title} {\bibinfo {title} {The kondo effect,
  conformal field theory and fusion rules},\ }\href
  {https://doi.org/10.1016/0550-3213(91)90419-X} {\bibfield  {journal}
  {\bibinfo  {journal} {Nuclear Physics B}\ }\textbf {\bibinfo {volume}
  {352}},\ \bibinfo {pages} {849} (\bibinfo {year} {1991})}\BibitemShut
  {NoStop}%
\bibitem [{\citenamefont {Potok}\ \emph {et~al.}(2007)\citenamefont {Potok},
  \citenamefont {Rau}, \citenamefont {Shtrikman}, \citenamefont {Oreg},\ and\
  \citenamefont {Goldhaber-Gordon}}]{potok2007}%
  \BibitemOpen
  \bibfield  {author} {\bibinfo {author} {\bibfnamefont {R.}~\bibnamefont
  {Potok}}, \bibinfo {author} {\bibfnamefont {I.}~\bibnamefont {Rau}}, \bibinfo
  {author} {\bibfnamefont {H.}~\bibnamefont {Shtrikman}}, \bibinfo {author}
  {\bibfnamefont {Y.}~\bibnamefont {Oreg}},\ and\ \bibinfo {author}
  {\bibfnamefont {D.}~\bibnamefont {Goldhaber-Gordon}},\ }\bibfield  {title}
  {\bibinfo {title} {Observation of the two-channel kondo effect},\ }\href
  {https://doi.org/10.1038/nature05556} {\bibfield  {journal} {\bibinfo
  {journal} {Nature}\ }\textbf {\bibinfo {volume} {446}},\ \bibinfo {pages}
  {167} (\bibinfo {year} {2007})}\BibitemShut {NoStop}%
\bibitem [{\citenamefont {Keller}\ \emph {et~al.}(2015)\citenamefont {Keller},
  \citenamefont {Peeters}, \citenamefont {Moca}, \citenamefont {Weymann},
  \citenamefont {Mahalu}, \citenamefont {Umansky}, \citenamefont {Zar{\'a}nd},\
  and\ \citenamefont {Goldhaber-Gordon}}]{keller2015}%
  \BibitemOpen
  \bibfield  {author} {\bibinfo {author} {\bibfnamefont {A.}~\bibnamefont
  {Keller}}, \bibinfo {author} {\bibfnamefont {L.}~\bibnamefont {Peeters}},
  \bibinfo {author} {\bibfnamefont {C.}~\bibnamefont {Moca}}, \bibinfo {author}
  {\bibfnamefont {I.}~\bibnamefont {Weymann}}, \bibinfo {author} {\bibfnamefont
  {D.}~\bibnamefont {Mahalu}}, \bibinfo {author} {\bibfnamefont
  {V.}~\bibnamefont {Umansky}}, \bibinfo {author} {\bibfnamefont
  {G.}~\bibnamefont {Zar{\'a}nd}},\ and\ \bibinfo {author} {\bibfnamefont
  {D.}~\bibnamefont {Goldhaber-Gordon}},\ }\bibfield  {title} {\bibinfo {title}
  {Universal fermi liquid crossover and quantum criticality in a mesoscopic
  system},\ }\href {https://doi.org/10.1038/nature15261} {\bibfield  {journal}
  {\bibinfo  {journal} {Nature}\ }\textbf {\bibinfo {volume} {526}},\ \bibinfo
  {pages} {237} (\bibinfo {year} {2015})}\BibitemShut {NoStop}%
\bibitem [{\citenamefont {Iftikhar}\ \emph {et~al.}(2015)\citenamefont
  {Iftikhar}, \citenamefont {Jezouin}, \citenamefont {Anthore}, \citenamefont
  {Gennser}, \citenamefont {Parmentier}, \citenamefont {Cavanna},\ and\
  \citenamefont {Pierre}}]{Iftikhar2015}%
  \BibitemOpen
  \bibfield  {author} {\bibinfo {author} {\bibfnamefont {Z.}~\bibnamefont
  {Iftikhar}}, \bibinfo {author} {\bibfnamefont {S.}~\bibnamefont {Jezouin}},
  \bibinfo {author} {\bibfnamefont {A.}~\bibnamefont {Anthore}}, \bibinfo
  {author} {\bibfnamefont {U.}~\bibnamefont {Gennser}}, \bibinfo {author}
  {\bibfnamefont {F.}~\bibnamefont {Parmentier}}, \bibinfo {author}
  {\bibfnamefont {A.}~\bibnamefont {Cavanna}},\ and\ \bibinfo {author}
  {\bibfnamefont {F.}~\bibnamefont {Pierre}},\ }\bibfield  {title} {\bibinfo
  {title} {Two-channel kondo effect and renormalization flow with macroscopic
  quantum charge states},\ }\href {https://doi.org/10.1038/nature15384}
  {\bibfield  {journal} {\bibinfo  {journal} {Nature}\ }\textbf {\bibinfo
  {volume} {526}},\ \bibinfo {pages} {233} (\bibinfo {year}
  {2015})}\BibitemShut {NoStop}%
\bibitem [{\citenamefont {Iftikhar}\ \emph {et~al.}(2018)\citenamefont
  {Iftikhar}, \citenamefont {Anthore}, \citenamefont {Mitchell}, \citenamefont
  {Parmentier}, \citenamefont {Gennser}, \citenamefont {Ouerghi}, \citenamefont
  {Cavanna}, \citenamefont {Mora}, \citenamefont {Simon},\ and\ \citenamefont
  {Pierre}}]{Iftikhar2018}%
  \BibitemOpen
  \bibfield  {author} {\bibinfo {author} {\bibfnamefont {Z.}~\bibnamefont
  {Iftikhar}}, \bibinfo {author} {\bibfnamefont {A.}~\bibnamefont {Anthore}},
  \bibinfo {author} {\bibfnamefont {A.}~\bibnamefont {Mitchell}}, \bibinfo
  {author} {\bibfnamefont {F.}~\bibnamefont {Parmentier}}, \bibinfo {author}
  {\bibfnamefont {U.}~\bibnamefont {Gennser}}, \bibinfo {author} {\bibfnamefont
  {A.}~\bibnamefont {Ouerghi}}, \bibinfo {author} {\bibfnamefont
  {A.}~\bibnamefont {Cavanna}}, \bibinfo {author} {\bibfnamefont
  {C.}~\bibnamefont {Mora}}, \bibinfo {author} {\bibfnamefont {P.}~\bibnamefont
  {Simon}},\ and\ \bibinfo {author} {\bibfnamefont {F.}~\bibnamefont
  {Pierre}},\ }\bibfield  {title} {\bibinfo {title} {Tunable quantum
  criticality and super-ballistic transport in a “charge” kondo circuit},\
  }\href {https://doi.org/10.1126/science.aan5592} {\bibfield  {journal}
  {\bibinfo  {journal} {Science}\ }\textbf {\bibinfo {volume} {360}},\ \bibinfo
  {pages} {1315} (\bibinfo {year} {2018})}\BibitemShut {NoStop}%
\bibitem [{\citenamefont {Karki}\ \emph {et~al.}(2023)\citenamefont {Karki},
  \citenamefont {Boulat}, \citenamefont {Pouse}, \citenamefont
  {Goldhaber-Gordon}, \citenamefont {Mitchell},\ and\ \citenamefont
  {Mora}}]{karski2023}%
  \BibitemOpen
  \bibfield  {author} {\bibinfo {author} {\bibfnamefont {D.}~\bibnamefont
  {Karki}}, \bibinfo {author} {\bibfnamefont {E.}~\bibnamefont {Boulat}},
  \bibinfo {author} {\bibfnamefont {W.}~\bibnamefont {Pouse}}, \bibinfo
  {author} {\bibfnamefont {D.}~\bibnamefont {Goldhaber-Gordon}}, \bibinfo
  {author} {\bibfnamefont {A.~K.}\ \bibnamefont {Mitchell}},\ and\ \bibinfo
  {author} {\bibfnamefont {C.}~\bibnamefont {Mora}},\ }\bibfield  {title}
  {\bibinfo {title} {Z 3 parafermion in the double charge kondo model},\ }\href
  {https://doi.org/10.1103/PhysRevB.107.L161108} {\bibfield  {journal}
  {\bibinfo  {journal} {Physical Review Letters}\ }\textbf {\bibinfo {volume}
  {130}},\ \bibinfo {pages} {146201} (\bibinfo {year} {2023})}\BibitemShut
  {NoStop}%
\bibitem [{\citenamefont {Lopes}\ \emph {et~al.}(2020)\citenamefont {Lopes},
  \citenamefont {Affleck},\ and\ \citenamefont {Sela}}]{Sela2020}%
  \BibitemOpen
  \bibfield  {author} {\bibinfo {author} {\bibfnamefont {P.~L.~S.}\
  \bibnamefont {Lopes}}, \bibinfo {author} {\bibfnamefont {I.}~\bibnamefont
  {Affleck}},\ and\ \bibinfo {author} {\bibfnamefont {E.}~\bibnamefont
  {Sela}},\ }\bibfield  {title} {\bibinfo {title} {Anyons in multichannel kondo
  systems},\ }\href {https://doi.org/10.1103/PhysRevB.101.085141} {\bibfield
  {journal} {\bibinfo  {journal} {Phys. Rev. B}\ }\textbf {\bibinfo {volume}
  {101}},\ \bibinfo {pages} {085141} (\bibinfo {year} {2020})}\BibitemShut
  {NoStop}%
\bibitem [{\citenamefont {Komijani}(2020)}]{Komijani2020}%
  \BibitemOpen
  \bibfield  {author} {\bibinfo {author} {\bibfnamefont {Y.}~\bibnamefont
  {Komijani}},\ }\bibfield  {title} {\bibinfo {title} {Isolating kondo anyons
  for topological quantum computation},\ }\href
  {https://doi.org/10.1103/PhysRevB.101.235131} {\bibfield  {journal} {\bibinfo
   {journal} {Phys. Rev. B}\ }\textbf {\bibinfo {volume} {101}},\ \bibinfo
  {pages} {235131} (\bibinfo {year} {2020})}\BibitemShut {NoStop}%
\bibitem [{\citenamefont {Ge}\ and\ \citenamefont {Komijani}(2022)}]{Ge2022}%
  \BibitemOpen
  \bibfield  {author} {\bibinfo {author} {\bibfnamefont {Y.}~\bibnamefont
  {Ge}}\ and\ \bibinfo {author} {\bibfnamefont {Y.}~\bibnamefont {Komijani}},\
  }\bibfield  {title} {\bibinfo {title} {Emergent spinon dispersion and
  symmetry breaking in two-channel kondo lattices},\ }\href
  {https://doi.org/10.1103/PhysRevLett.129.077202} {\bibfield  {journal}
  {\bibinfo  {journal} {Phys. Rev. Lett.}\ }\textbf {\bibinfo {volume} {129}},\
  \bibinfo {pages} {077202} (\bibinfo {year} {2022})}\BibitemShut {NoStop}%
\bibitem [{\citenamefont {B\'eri}\ and\ \citenamefont
  {Cooper}(2012)}]{BeriCooper2012}%
  \BibitemOpen
  \bibfield  {author} {\bibinfo {author} {\bibfnamefont {B.}~\bibnamefont
  {B\'eri}}\ and\ \bibinfo {author} {\bibfnamefont {N.~R.}\ \bibnamefont
  {Cooper}},\ }\bibfield  {title} {\bibinfo {title} {Topological kondo effect
  with majorana fermions},\ }\href
  {https://doi.org/10.1103/PhysRevLett.109.156803} {\bibfield  {journal}
  {\bibinfo  {journal} {Phys. Rev. Lett.}\ }\textbf {\bibinfo {volume} {109}},\
  \bibinfo {pages} {156803} (\bibinfo {year} {2012})}\BibitemShut {NoStop}%
\bibitem [{\citenamefont {Altland}\ and\ \citenamefont
  {Egger}(2013)}]{TopKondo}%
  \BibitemOpen
  \bibfield  {author} {\bibinfo {author} {\bibfnamefont {A.}~\bibnamefont
  {Altland}}\ and\ \bibinfo {author} {\bibfnamefont {R.}~\bibnamefont
  {Egger}},\ }\bibfield  {title} {\bibinfo {title} {Multiterminal
  coulomb-majorana junction},\ }\href
  {https://doi.org/10.1103/PhysRevLett.110.196401} {\bibfield  {journal}
  {\bibinfo  {journal} {Phys. Rev. Lett.}\ }\textbf {\bibinfo {volume} {110}},\
  \bibinfo {pages} {196401} (\bibinfo {year} {2013})}\BibitemShut {NoStop}%
\bibitem [{\citenamefont {Altland}\ \emph {et~al.}(2014)\citenamefont
  {Altland}, \citenamefont {B\'eri}, \citenamefont {Egger},\ and\ \citenamefont
  {Tsvelik}}]{TopKondoWe}%
  \BibitemOpen
  \bibfield  {author} {\bibinfo {author} {\bibfnamefont {A.}~\bibnamefont
  {Altland}}, \bibinfo {author} {\bibfnamefont {B.}~\bibnamefont {B\'eri}},
  \bibinfo {author} {\bibfnamefont {R.}~\bibnamefont {Egger}},\ and\ \bibinfo
  {author} {\bibfnamefont {A.~M.}\ \bibnamefont {Tsvelik}},\ }\bibfield
  {title} {\bibinfo {title} {Multichannel kondo impurity dynamics in a majorana
  device},\ }\href {https://doi.org/10.1103/PhysRevLett.113.076401} {\bibfield
  {journal} {\bibinfo  {journal} {Phys. Rev. Lett.}\ }\textbf {\bibinfo
  {volume} {113}},\ \bibinfo {pages} {076401} (\bibinfo {year}
  {2014})}\BibitemShut {NoStop}%
\bibitem [{\citenamefont {V\"ayrynen}\ \emph {et~al.}(2020)\citenamefont
  {V\"ayrynen}, \citenamefont {Feiguin},\ and\ \citenamefont
  {Lutchyn}}]{Roman2020}%
  \BibitemOpen
  \bibfield  {author} {\bibinfo {author} {\bibfnamefont {J.~I.}\ \bibnamefont
  {V\"ayrynen}}, \bibinfo {author} {\bibfnamefont {A.~E.}\ \bibnamefont
  {Feiguin}},\ and\ \bibinfo {author} {\bibfnamefont {R.~M.}\ \bibnamefont
  {Lutchyn}},\ }\bibfield  {title} {\bibinfo {title} {Signatures of topological
  ground state degeneracy in majorana islands},\ }\href
  {https://doi.org/10.1103/PhysRevResearch.2.043228} {\bibfield  {journal}
  {\bibinfo  {journal} {Phys. Rev. Res.}\ }\textbf {\bibinfo {volume} {2}},\
  \bibinfo {pages} {043228} (\bibinfo {year} {2020})}\BibitemShut {NoStop}%
\bibitem [{\citenamefont {Liu}\ \emph {et~al.}(2021)\citenamefont {Liu},
  \citenamefont {Cao}, \citenamefont {Liu}, \citenamefont {Zhang},\ and\
  \citenamefont {Liu}}]{Dong2021}%
  \BibitemOpen
  \bibfield  {author} {\bibinfo {author} {\bibfnamefont {D.}~\bibnamefont
  {Liu}}, \bibinfo {author} {\bibfnamefont {Z.}~\bibnamefont {Cao}}, \bibinfo
  {author} {\bibfnamefont {X.}~\bibnamefont {Liu}}, \bibinfo {author}
  {\bibfnamefont {H.}~\bibnamefont {Zhang}},\ and\ \bibinfo {author}
  {\bibfnamefont {D.~E.}\ \bibnamefont {Liu}},\ }\bibfield  {title} {\bibinfo
  {title} {Topological kondo device for distinguishing quasi-majorana and
  majorana signatures},\ }\href {https://doi.org/10.1103/PhysRevB.104.205125}
  {\bibfield  {journal} {\bibinfo  {journal} {Phys. Rev. B}\ }\textbf {\bibinfo
  {volume} {104}},\ \bibinfo {pages} {205125} (\bibinfo {year}
  {2021})}\BibitemShut {NoStop}%
\bibitem [{\citenamefont {Beenakker}(2013)}]{Beenakker2013}%
  \BibitemOpen
  \bibfield  {author} {\bibinfo {author} {\bibfnamefont {C.}~\bibnamefont
  {Beenakker}},\ }\bibfield  {title} {\bibinfo {title} {Search for majorana
  fermions in superconductors},\ }\href
  {https://doi.org/10.1146/annurev-conmatphys-030212-184337} {\bibfield
  {journal} {\bibinfo  {journal} {Annu. Rev. Condens. Matter Phys.}\ }\textbf
  {\bibinfo {volume} {4}},\ \bibinfo {pages} {113} (\bibinfo {year}
  {2013})}\BibitemShut {NoStop}%
\bibitem [{\citenamefont {Sato}\ and\ \citenamefont
  {Ando}(2017)}]{Sato2017topo}%
  \BibitemOpen
  \bibfield  {author} {\bibinfo {author} {\bibfnamefont {M.}~\bibnamefont
  {Sato}}\ and\ \bibinfo {author} {\bibfnamefont {Y.}~\bibnamefont {Ando}},\
  }\bibfield  {title} {\bibinfo {title} {Topological superconductors: a
  review},\ }\href {https://doi.org/10.1088/1361-6633/aa6ac7} {\bibfield
  {journal} {\bibinfo  {journal} {Reports on Progress in Physics}\ }\textbf
  {\bibinfo {volume} {80}},\ \bibinfo {pages} {076501} (\bibinfo {year}
  {2017})}\BibitemShut {NoStop}%
\bibitem [{\citenamefont {Xu}\ \emph {et~al.}(2016)\citenamefont {Xu},
  \citenamefont {Lian}, \citenamefont {Tang}, \citenamefont {Qi},\ and\
  \citenamefont {Zhang}}]{Zhang2016}%
  \BibitemOpen
  \bibfield  {author} {\bibinfo {author} {\bibfnamefont {G.}~\bibnamefont
  {Xu}}, \bibinfo {author} {\bibfnamefont {B.}~\bibnamefont {Lian}}, \bibinfo
  {author} {\bibfnamefont {P.}~\bibnamefont {Tang}}, \bibinfo {author}
  {\bibfnamefont {X.-L.}\ \bibnamefont {Qi}},\ and\ \bibinfo {author}
  {\bibfnamefont {S.-C.}\ \bibnamefont {Zhang}},\ }\bibfield  {title} {\bibinfo
  {title} {Topological superconductivity on the surface of fe-based
  superconductors},\ }\href {https://doi.org/10.1103/PhysRevLett.117.047001}
  {\bibfield  {journal} {\bibinfo  {journal} {Phys. Rev. Lett.}\ }\textbf
  {\bibinfo {volume} {117}},\ \bibinfo {pages} {047001} (\bibinfo {year}
  {2016})}\BibitemShut {NoStop}%
\bibitem [{\citenamefont {Wang}\ \emph {et~al.}(2018)\citenamefont {Wang},
  \citenamefont {Kong}, \citenamefont {Fan}, \citenamefont {Chen},
  \citenamefont {Zhu}, \citenamefont {Liu}, \citenamefont {Cao}, \citenamefont
  {Sun}, \citenamefont {Du}, \citenamefont {Schneeloch}, \citenamefont {Zhong},
  \citenamefont {Gu}, \citenamefont {Fu}, \citenamefont {Ding},\ and\
  \citenamefont {Gao}}]{Wang2018}%
  \BibitemOpen
  \bibfield  {author} {\bibinfo {author} {\bibfnamefont {D.}~\bibnamefont
  {Wang}}, \bibinfo {author} {\bibfnamefont {L.}~\bibnamefont {Kong}}, \bibinfo
  {author} {\bibfnamefont {P.}~\bibnamefont {Fan}}, \bibinfo {author}
  {\bibfnamefont {H.}~\bibnamefont {Chen}}, \bibinfo {author} {\bibfnamefont
  {S.}~\bibnamefont {Zhu}}, \bibinfo {author} {\bibfnamefont {W.}~\bibnamefont
  {Liu}}, \bibinfo {author} {\bibfnamefont {L.}~\bibnamefont {Cao}}, \bibinfo
  {author} {\bibfnamefont {Y.}~\bibnamefont {Sun}}, \bibinfo {author}
  {\bibfnamefont {S.}~\bibnamefont {Du}}, \bibinfo {author} {\bibfnamefont
  {J.}~\bibnamefont {Schneeloch}}, \bibinfo {author} {\bibfnamefont
  {R.}~\bibnamefont {Zhong}}, \bibinfo {author} {\bibfnamefont
  {G.}~\bibnamefont {Gu}}, \bibinfo {author} {\bibfnamefont {L.}~\bibnamefont
  {Fu}}, \bibinfo {author} {\bibfnamefont {H.}~\bibnamefont {Ding}},\ and\
  \bibinfo {author} {\bibfnamefont {H.-J.}\ \bibnamefont {Gao}},\ }\bibfield
  {title} {\bibinfo {title} {Evidence for majorana bound states in an
  iron-based superconductor},\ }\href {https://doi.org/10.1126/science.aao1797}
  {\bibfield  {journal} {\bibinfo  {journal} {Science}\ }\textbf {\bibinfo
  {volume} {362}},\ \bibinfo {pages} {333} (\bibinfo {year}
  {2018})}\BibitemShut {NoStop}%
\bibitem [{\citenamefont {Zhang}\ \emph {et~al.}(2018)\citenamefont {Zhang},
  \citenamefont {Yaji}, \citenamefont {Hashimoto}, \citenamefont {Ota},
  \citenamefont {Kondo}, \citenamefont {Okazaki}, \citenamefont {Wang},
  \citenamefont {Wen}, \citenamefont {Gu}, \citenamefont {Ding} \emph
  {et~al.}}]{Zhang2018}%
  \BibitemOpen
  \bibfield  {author} {\bibinfo {author} {\bibfnamefont {P.}~\bibnamefont
  {Zhang}}, \bibinfo {author} {\bibfnamefont {K.}~\bibnamefont {Yaji}},
  \bibinfo {author} {\bibfnamefont {T.}~\bibnamefont {Hashimoto}}, \bibinfo
  {author} {\bibfnamefont {Y.}~\bibnamefont {Ota}}, \bibinfo {author}
  {\bibfnamefont {T.}~\bibnamefont {Kondo}}, \bibinfo {author} {\bibfnamefont
  {K.}~\bibnamefont {Okazaki}}, \bibinfo {author} {\bibfnamefont
  {Z.}~\bibnamefont {Wang}}, \bibinfo {author} {\bibfnamefont {J.}~\bibnamefont
  {Wen}}, \bibinfo {author} {\bibfnamefont {G.}~\bibnamefont {Gu}}, \bibinfo
  {author} {\bibfnamefont {H.}~\bibnamefont {Ding}}, \emph {et~al.},\
  }\bibfield  {title} {\bibinfo {title} {Observation of topological
  superconductivity on the surface of an iron-based superconductor},\ }\href
  {https://doi.org/10.1126/science.aan4596} {\bibfield  {journal} {\bibinfo
  {journal} {Science}\ }\textbf {\bibinfo {volume} {360}},\ \bibinfo {pages}
  {182} (\bibinfo {year} {2018})}\BibitemShut {NoStop}%
\bibitem [{\citenamefont {Machida}\ \emph {et~al.}(2019)\citenamefont
  {Machida}, \citenamefont {Sun}, \citenamefont {Pyon}, \citenamefont {Takeda},
  \citenamefont {Kohsaka}, \citenamefont {Hanaguri}, \citenamefont {Sasagawa},\
  and\ \citenamefont {Tamegai}}]{Machida2018}%
  \BibitemOpen
  \bibfield  {author} {\bibinfo {author} {\bibfnamefont {T.}~\bibnamefont
  {Machida}}, \bibinfo {author} {\bibfnamefont {Y.}~\bibnamefont {Sun}},
  \bibinfo {author} {\bibfnamefont {S.}~\bibnamefont {Pyon}}, \bibinfo {author}
  {\bibfnamefont {S.}~\bibnamefont {Takeda}}, \bibinfo {author} {\bibfnamefont
  {Y.}~\bibnamefont {Kohsaka}}, \bibinfo {author} {\bibfnamefont
  {T.}~\bibnamefont {Hanaguri}}, \bibinfo {author} {\bibfnamefont
  {T.}~\bibnamefont {Sasagawa}},\ and\ \bibinfo {author} {\bibfnamefont
  {T.}~\bibnamefont {Tamegai}},\ }\bibfield  {title} {\bibinfo {title}
  {Zero-energy vortex bound state in the superconducting topological surface
  state of fe (se, te)},\ }\href {https://doi.org/10.1038/s41563-019-0397-1}
  {\bibfield  {journal} {\bibinfo  {journal} {Nature materials}\ }\textbf
  {\bibinfo {volume} {18}},\ \bibinfo {pages} {811} (\bibinfo {year}
  {2019})}\BibitemShut {NoStop}%
\bibitem [{\citenamefont {Liu}\ \emph {et~al.}(2018)\citenamefont {Liu},
  \citenamefont {Chen}, \citenamefont {Zhang}, \citenamefont {Peng},
  \citenamefont {Yan}, \citenamefont {Wen}, \citenamefont {Lou}, \citenamefont
  {Huang}, \citenamefont {Tian}, \citenamefont {Dong}, \citenamefont {Wang},
  \citenamefont {Bao}, \citenamefont {Wang}, \citenamefont {Yin}, \citenamefont
  {Zhao},\ and\ \citenamefont {Feng}}]{Liu2018}%
  \BibitemOpen
  \bibfield  {author} {\bibinfo {author} {\bibfnamefont {Q.}~\bibnamefont
  {Liu}}, \bibinfo {author} {\bibfnamefont {C.}~\bibnamefont {Chen}}, \bibinfo
  {author} {\bibfnamefont {T.}~\bibnamefont {Zhang}}, \bibinfo {author}
  {\bibfnamefont {R.}~\bibnamefont {Peng}}, \bibinfo {author} {\bibfnamefont
  {Y.-J.}\ \bibnamefont {Yan}}, \bibinfo {author} {\bibfnamefont {C.-H.-P.}\
  \bibnamefont {Wen}}, \bibinfo {author} {\bibfnamefont {X.}~\bibnamefont
  {Lou}}, \bibinfo {author} {\bibfnamefont {Y.-L.}\ \bibnamefont {Huang}},
  \bibinfo {author} {\bibfnamefont {J.-P.}\ \bibnamefont {Tian}}, \bibinfo
  {author} {\bibfnamefont {X.-L.}\ \bibnamefont {Dong}}, \bibinfo {author}
  {\bibfnamefont {G.-W.}\ \bibnamefont {Wang}}, \bibinfo {author}
  {\bibfnamefont {W.-C.}\ \bibnamefont {Bao}}, \bibinfo {author} {\bibfnamefont
  {Q.-H.}\ \bibnamefont {Wang}}, \bibinfo {author} {\bibfnamefont {Z.-P.}\
  \bibnamefont {Yin}}, \bibinfo {author} {\bibfnamefont {Z.-X.}\ \bibnamefont
  {Zhao}},\ and\ \bibinfo {author} {\bibfnamefont {D.-L.}\ \bibnamefont
  {Feng}},\ }\bibfield  {title} {\bibinfo {title} {Robust and clean majorana
  zero mode in the vortex core of high-temperature superconductor
  $\mathbf{(}{\mathrm{li}}_{0.84}{\mathrm{fe}}_{0.16}\mathbf{)}\mathrm{OHFeSe}$},\
  }\href {https://doi.org/10.1103/PhysRevX.8.041056} {\bibfield  {journal}
  {\bibinfo  {journal} {Phys. Rev. X}\ }\textbf {\bibinfo {volume} {8}},\
  \bibinfo {pages} {041056} (\bibinfo {year} {2018})}\BibitemShut {NoStop}%
\bibitem [{\citenamefont {Liu}\ \emph {et~al.}(2020{\natexlab{a}})\citenamefont
  {Liu}, \citenamefont {Cao}, \citenamefont {Zhu}, \citenamefont {Kong},
  \citenamefont {Wang}, \citenamefont {Papaj}, \citenamefont {Zhang},
  \citenamefont {Liu}, \citenamefont {Chen}, \citenamefont {Li} \emph
  {et~al.}}]{Liu2020}%
  \BibitemOpen
  \bibfield  {author} {\bibinfo {author} {\bibfnamefont {W.}~\bibnamefont
  {Liu}}, \bibinfo {author} {\bibfnamefont {L.}~\bibnamefont {Cao}}, \bibinfo
  {author} {\bibfnamefont {S.}~\bibnamefont {Zhu}}, \bibinfo {author}
  {\bibfnamefont {L.}~\bibnamefont {Kong}}, \bibinfo {author} {\bibfnamefont
  {G.}~\bibnamefont {Wang}}, \bibinfo {author} {\bibfnamefont {M.}~\bibnamefont
  {Papaj}}, \bibinfo {author} {\bibfnamefont {P.}~\bibnamefont {Zhang}},
  \bibinfo {author} {\bibfnamefont {Y.-B.}\ \bibnamefont {Liu}}, \bibinfo
  {author} {\bibfnamefont {H.}~\bibnamefont {Chen}}, \bibinfo {author}
  {\bibfnamefont {G.}~\bibnamefont {Li}}, \emph {et~al.},\ }\bibfield  {title}
  {\bibinfo {title} {A new majorana platform in an fe-as bilayer
  superconductor},\ }\href {https://doi.org/10.1038/s41467-020-19487-1}
  {\bibfield  {journal} {\bibinfo  {journal} {Nature Communications}\ }\textbf
  {\bibinfo {volume} {11}},\ \bibinfo {pages} {5688} (\bibinfo {year}
  {2020}{\natexlab{a}})}\BibitemShut {NoStop}%
\bibitem [{\citenamefont {Yin}\ \emph {et~al.}(2015)\citenamefont {Yin},
  \citenamefont {Wu}, \citenamefont {Wang}, \citenamefont {Ye}, \citenamefont
  {Gong}, \citenamefont {Hou}, \citenamefont {Shan}, \citenamefont {Li},
  \citenamefont {Liang}, \citenamefont {Wu} \emph {et~al.}}]{Yin2015}%
  \BibitemOpen
  \bibfield  {author} {\bibinfo {author} {\bibfnamefont {J.-X.}\ \bibnamefont
  {Yin}}, \bibinfo {author} {\bibfnamefont {Z.}~\bibnamefont {Wu}}, \bibinfo
  {author} {\bibfnamefont {J.}~\bibnamefont {Wang}}, \bibinfo {author}
  {\bibfnamefont {Z.}~\bibnamefont {Ye}}, \bibinfo {author} {\bibfnamefont
  {J.}~\bibnamefont {Gong}}, \bibinfo {author} {\bibfnamefont {X.}~\bibnamefont
  {Hou}}, \bibinfo {author} {\bibfnamefont {L.}~\bibnamefont {Shan}}, \bibinfo
  {author} {\bibfnamefont {A.}~\bibnamefont {Li}}, \bibinfo {author}
  {\bibfnamefont {X.}~\bibnamefont {Liang}}, \bibinfo {author} {\bibfnamefont
  {X.}~\bibnamefont {Wu}}, \emph {et~al.},\ }\bibfield  {title} {\bibinfo
  {title} {Observation of a robust zero-energy bound state in iron-based
  superconductor fe (te, se)},\ }\href {https://doi.org/10.1038/nphys3371}
  {\bibfield  {journal} {\bibinfo  {journal} {Nature Physics}\ }\textbf
  {\bibinfo {volume} {11}},\ \bibinfo {pages} {543} (\bibinfo {year}
  {2015})}\BibitemShut {NoStop}%
\bibitem [{\citenamefont {Zhu}\ \emph {et~al.}(2020)\citenamefont {Zhu},
  \citenamefont {Kong}, \citenamefont {Cao}, \citenamefont {Chen},
  \citenamefont {Papaj}, \citenamefont {Du}, \citenamefont {Xing},
  \citenamefont {Liu}, \citenamefont {Wang}, \citenamefont {Shen} \emph
  {et~al.}}]{Zhu2020}%
  \BibitemOpen
  \bibfield  {author} {\bibinfo {author} {\bibfnamefont {S.}~\bibnamefont
  {Zhu}}, \bibinfo {author} {\bibfnamefont {L.}~\bibnamefont {Kong}}, \bibinfo
  {author} {\bibfnamefont {L.}~\bibnamefont {Cao}}, \bibinfo {author}
  {\bibfnamefont {H.}~\bibnamefont {Chen}}, \bibinfo {author} {\bibfnamefont
  {M.}~\bibnamefont {Papaj}}, \bibinfo {author} {\bibfnamefont
  {S.}~\bibnamefont {Du}}, \bibinfo {author} {\bibfnamefont {Y.}~\bibnamefont
  {Xing}}, \bibinfo {author} {\bibfnamefont {W.}~\bibnamefont {Liu}}, \bibinfo
  {author} {\bibfnamefont {D.}~\bibnamefont {Wang}}, \bibinfo {author}
  {\bibfnamefont {C.}~\bibnamefont {Shen}}, \emph {et~al.},\ }\bibfield
  {title} {\bibinfo {title} {Nearly quantized conductance plateau of vortex
  zero mode in an iron-based superconductor},\ }\href
  {https://doi.org/10.1126/science.aax0274} {\bibfield  {journal} {\bibinfo
  {journal} {Science}\ }\textbf {\bibinfo {volume} {367}},\ \bibinfo {pages}
  {189} (\bibinfo {year} {2020})}\BibitemShut {NoStop}%
\bibitem [{\citenamefont {Alicea}(2012)}]{Alicea2012}%
  \BibitemOpen
  \bibfield  {author} {\bibinfo {author} {\bibfnamefont {J.}~\bibnamefont
  {Alicea}},\ }\bibfield  {title} {\bibinfo {title} {New directions in the
  pursuit of majorana fermions in solid state systems},\ }\href
  {https://doi.org/10.1088/0034-4885/75/7/076501} {\bibfield  {journal}
  {\bibinfo  {journal} {Reports on progress in physics}\ }\textbf {\bibinfo
  {volume} {75}},\ \bibinfo {pages} {076501} (\bibinfo {year}
  {2012})}\BibitemShut {NoStop}%
\bibitem [{\citenamefont {Caroli}\ \emph {et~al.}(1964)\citenamefont {Caroli},
  \citenamefont {{De Gennes}},\ and\ \citenamefont {Matricon}}]{CAROLI1964}%
  \BibitemOpen
  \bibfield  {author} {\bibinfo {author} {\bibfnamefont {C.}~\bibnamefont
  {Caroli}}, \bibinfo {author} {\bibfnamefont {P.}~\bibnamefont {{De
  Gennes}}},\ and\ \bibinfo {author} {\bibfnamefont {J.}~\bibnamefont
  {Matricon}},\ }\bibfield  {title} {\bibinfo {title} {Bound fermion states on
  a vortex line in a type ii superconductor},\ }\href
  {https://doi.org/https://doi.org/10.1016/0031-9163(64)90375-0} {\bibfield
  {journal} {\bibinfo  {journal} {Physics Letters}\ }\textbf {\bibinfo {volume}
  {9}},\ \bibinfo {pages} {307} (\bibinfo {year} {1964})}\BibitemShut {NoStop}%
\bibitem [{\citenamefont {Kong}\ \emph {et~al.}(2019)\citenamefont {Kong},
  \citenamefont {Zhu}, \citenamefont {Papaj}, \citenamefont {Chen},
  \citenamefont {Cao}, \citenamefont {Isobe}, \citenamefont {Xing},
  \citenamefont {Liu}, \citenamefont {Wang}, \citenamefont {Fan} \emph
  {et~al.}}]{kong2019half}%
  \BibitemOpen
  \bibfield  {author} {\bibinfo {author} {\bibfnamefont {L.}~\bibnamefont
  {Kong}}, \bibinfo {author} {\bibfnamefont {S.}~\bibnamefont {Zhu}}, \bibinfo
  {author} {\bibfnamefont {M.}~\bibnamefont {Papaj}}, \bibinfo {author}
  {\bibfnamefont {H.}~\bibnamefont {Chen}}, \bibinfo {author} {\bibfnamefont
  {L.}~\bibnamefont {Cao}}, \bibinfo {author} {\bibfnamefont {H.}~\bibnamefont
  {Isobe}}, \bibinfo {author} {\bibfnamefont {Y.}~\bibnamefont {Xing}},
  \bibinfo {author} {\bibfnamefont {W.}~\bibnamefont {Liu}}, \bibinfo {author}
  {\bibfnamefont {D.}~\bibnamefont {Wang}}, \bibinfo {author} {\bibfnamefont
  {P.}~\bibnamefont {Fan}}, \emph {et~al.},\ }\bibfield  {title} {\bibinfo
  {title} {Half-integer level shift of vortex bound states in an iron-based
  superconductor},\ }\href
  {https://doi.org/https://doi.org/10.1038/s41567-019-0630-5} {\bibfield
  {journal} {\bibinfo  {journal} {Nature Physics}\ }\textbf {\bibinfo {volume}
  {15}},\ \bibinfo {pages} {1181} (\bibinfo {year} {2019})}\BibitemShut
  {NoStop}%
\bibitem [{\citenamefont {Jiang}\ \emph {et~al.}(2019)\citenamefont {Jiang},
  \citenamefont {Dai},\ and\ \citenamefont {Wang}}]{Jiang2019}%
  \BibitemOpen
  \bibfield  {author} {\bibinfo {author} {\bibfnamefont {K.}~\bibnamefont
  {Jiang}}, \bibinfo {author} {\bibfnamefont {X.}~\bibnamefont {Dai}},\ and\
  \bibinfo {author} {\bibfnamefont {Z.}~\bibnamefont {Wang}},\ }\bibfield
  {title} {\bibinfo {title} {Quantum anomalous vortex and majorana zero mode in
  iron-based superconductor fe(te,se)},\ }\href
  {https://doi.org/10.1103/PhysRevX.9.011033} {\bibfield  {journal} {\bibinfo
  {journal} {Phys. Rev. X}\ }\textbf {\bibinfo {volume} {9}},\ \bibinfo {pages}
  {011033} (\bibinfo {year} {2019})}\BibitemShut {NoStop}%
\bibitem [{\citenamefont {Zhang}\ \emph {et~al.}(2020)\citenamefont {Zhang},
  \citenamefont {Yin}, \citenamefont {Dai}, \citenamefont {Zhao}, \citenamefont
  {Chang}, \citenamefont {Shumiya}, \citenamefont {Jiang}, \citenamefont
  {Zheng}, \citenamefont {Bian}, \citenamefont {Multer}, \citenamefont
  {Litskevich}, \citenamefont {Chang}, \citenamefont {Belopolski},
  \citenamefont {Cochran}, \citenamefont {Wu}, \citenamefont {Wu},
  \citenamefont {Luo}, \citenamefont {Chen}, \citenamefont {Lin}, \citenamefont
  {Chou}, \citenamefont {Wang}, \citenamefont {Jin}, \citenamefont {Sankar},
  \citenamefont {Wang},\ and\ \citenamefont {Hasan}}]{Zahid2020}%
  \BibitemOpen
  \bibfield  {author} {\bibinfo {author} {\bibfnamefont {S.~S.}\ \bibnamefont
  {Zhang}}, \bibinfo {author} {\bibfnamefont {J.-X.}\ \bibnamefont {Yin}},
  \bibinfo {author} {\bibfnamefont {G.}~\bibnamefont {Dai}}, \bibinfo {author}
  {\bibfnamefont {L.}~\bibnamefont {Zhao}}, \bibinfo {author} {\bibfnamefont
  {T.-R.}\ \bibnamefont {Chang}}, \bibinfo {author} {\bibfnamefont
  {N.}~\bibnamefont {Shumiya}}, \bibinfo {author} {\bibfnamefont
  {K.}~\bibnamefont {Jiang}}, \bibinfo {author} {\bibfnamefont
  {H.}~\bibnamefont {Zheng}}, \bibinfo {author} {\bibfnamefont
  {G.}~\bibnamefont {Bian}}, \bibinfo {author} {\bibfnamefont {D.}~\bibnamefont
  {Multer}}, \bibinfo {author} {\bibfnamefont {M.}~\bibnamefont {Litskevich}},
  \bibinfo {author} {\bibfnamefont {G.}~\bibnamefont {Chang}}, \bibinfo
  {author} {\bibfnamefont {I.}~\bibnamefont {Belopolski}}, \bibinfo {author}
  {\bibfnamefont {T.~A.}\ \bibnamefont {Cochran}}, \bibinfo {author}
  {\bibfnamefont {X.}~\bibnamefont {Wu}}, \bibinfo {author} {\bibfnamefont
  {D.}~\bibnamefont {Wu}}, \bibinfo {author} {\bibfnamefont {J.}~\bibnamefont
  {Luo}}, \bibinfo {author} {\bibfnamefont {G.}~\bibnamefont {Chen}}, \bibinfo
  {author} {\bibfnamefont {H.}~\bibnamefont {Lin}}, \bibinfo {author}
  {\bibfnamefont {F.-C.}\ \bibnamefont {Chou}}, \bibinfo {author}
  {\bibfnamefont {X.}~\bibnamefont {Wang}}, \bibinfo {author} {\bibfnamefont
  {C.}~\bibnamefont {Jin}}, \bibinfo {author} {\bibfnamefont {R.}~\bibnamefont
  {Sankar}}, \bibinfo {author} {\bibfnamefont {Z.}~\bibnamefont {Wang}},\ and\
  \bibinfo {author} {\bibfnamefont {M.~Z.}\ \bibnamefont {Hasan}},\ }\bibfield
  {title} {\bibinfo {title} {Field-free platform for majorana-like zero mode in
  superconductors with a topological surface state},\ }\href
  {https://doi.org/10.1103/PhysRevB.101.100507} {\bibfield  {journal} {\bibinfo
   {journal} {Phys. Rev. B}\ }\textbf {\bibinfo {volume} {101}},\ \bibinfo
  {pages} {100507} (\bibinfo {year} {2020})}\BibitemShut {NoStop}%
\bibitem [{\citenamefont {Liu}\ \emph {et~al.}(2020{\natexlab{b}})\citenamefont
  {Liu}, \citenamefont {Chen}, \citenamefont {Liu}, \citenamefont {Wang},
  \citenamefont {Liu}, \citenamefont {Ye}, \citenamefont {Wang}, \citenamefont
  {Hu},\ and\ \citenamefont {Wang}}]{liu2020zero}%
  \BibitemOpen
  \bibfield  {author} {\bibinfo {author} {\bibfnamefont {C.}~\bibnamefont
  {Liu}}, \bibinfo {author} {\bibfnamefont {C.}~\bibnamefont {Chen}}, \bibinfo
  {author} {\bibfnamefont {X.}~\bibnamefont {Liu}}, \bibinfo {author}
  {\bibfnamefont {Z.}~\bibnamefont {Wang}}, \bibinfo {author} {\bibfnamefont
  {Y.}~\bibnamefont {Liu}}, \bibinfo {author} {\bibfnamefont {S.}~\bibnamefont
  {Ye}}, \bibinfo {author} {\bibfnamefont {Z.}~\bibnamefont {Wang}}, \bibinfo
  {author} {\bibfnamefont {J.}~\bibnamefont {Hu}},\ and\ \bibinfo {author}
  {\bibfnamefont {J.}~\bibnamefont {Wang}},\ }\bibfield  {title} {\bibinfo
  {title} {Zero-energy bound states in the high-temperature superconductors at
  the two-dimensional limit},\ }\href {https://doi.org/10.1126/sciadv.aax7547}
  {\bibfield  {journal} {\bibinfo  {journal} {Science advances}\ }\textbf
  {\bibinfo {volume} {6}},\ \bibinfo {pages} {eaax7547} (\bibinfo {year}
  {2020}{\natexlab{b}})}\BibitemShut {NoStop}%
\bibitem [{\citenamefont {Fan}\ \emph {et~al.}(2021)\citenamefont {Fan},
  \citenamefont {Yang}, \citenamefont {Qian}, \citenamefont {Chen},
  \citenamefont {Zhang}, \citenamefont {Li}, \citenamefont {Huang},
  \citenamefont {Xing}, \citenamefont {Kong}, \citenamefont {Liu} \emph
  {et~al.}}]{Fan2021}%
  \BibitemOpen
  \bibfield  {author} {\bibinfo {author} {\bibfnamefont {P.}~\bibnamefont
  {Fan}}, \bibinfo {author} {\bibfnamefont {F.}~\bibnamefont {Yang}}, \bibinfo
  {author} {\bibfnamefont {G.}~\bibnamefont {Qian}}, \bibinfo {author}
  {\bibfnamefont {H.}~\bibnamefont {Chen}}, \bibinfo {author} {\bibfnamefont
  {Y.-Y.}\ \bibnamefont {Zhang}}, \bibinfo {author} {\bibfnamefont
  {G.}~\bibnamefont {Li}}, \bibinfo {author} {\bibfnamefont {Z.}~\bibnamefont
  {Huang}}, \bibinfo {author} {\bibfnamefont {Y.}~\bibnamefont {Xing}},
  \bibinfo {author} {\bibfnamefont {L.}~\bibnamefont {Kong}}, \bibinfo {author}
  {\bibfnamefont {W.}~\bibnamefont {Liu}}, \emph {et~al.},\ }\bibfield  {title}
  {\bibinfo {title} {Observation of magnetic adatom-induced majorana vortex and
  its hybridization with field-induced majorana vortex in an iron-based
  superconductor},\ }\href
  {https://doi.org/https://doi.org/10.1038/s41467-021-21646-x} {\bibfield
  {journal} {\bibinfo  {journal} {Nature communications}\ }\textbf {\bibinfo
  {volume} {12}},\ \bibinfo {pages} {1348} (\bibinfo {year}
  {2021})}\BibitemShut {NoStop}%
\bibitem [{\citenamefont {Chen}\ \emph {et~al.}(2020)\citenamefont {Chen},
  \citenamefont {Jiang}, \citenamefont {Zhang}, \citenamefont {Liu},
  \citenamefont {Liu}, \citenamefont {Wang},\ and\ \citenamefont
  {Wang}}]{Chen2020}%
  \BibitemOpen
  \bibfield  {author} {\bibinfo {author} {\bibfnamefont {C.}~\bibnamefont
  {Chen}}, \bibinfo {author} {\bibfnamefont {K.}~\bibnamefont {Jiang}},
  \bibinfo {author} {\bibfnamefont {Y.}~\bibnamefont {Zhang}}, \bibinfo
  {author} {\bibfnamefont {C.}~\bibnamefont {Liu}}, \bibinfo {author}
  {\bibfnamefont {Y.}~\bibnamefont {Liu}}, \bibinfo {author} {\bibfnamefont
  {Z.}~\bibnamefont {Wang}},\ and\ \bibinfo {author} {\bibfnamefont
  {J.}~\bibnamefont {Wang}},\ }\bibfield  {title} {\bibinfo {title} {Atomic
  line defects and zero-energy end states in monolayer fe (te, se)
  high-temperature superconductors},\ }\href
  {https://doi.org/10.1038/s41567-020-0813-0} {\bibfield  {journal} {\bibinfo
  {journal} {Nature Physics}\ }\textbf {\bibinfo {volume} {16}},\ \bibinfo
  {pages} {536} (\bibinfo {year} {2020})}\BibitemShut {NoStop}%
\bibitem [{\citenamefont {Wang}\ \emph {et~al.}(2020)\citenamefont {Wang},
  \citenamefont {Rodriguez}, \citenamefont {Jiao}, \citenamefont {Howard},
  \citenamefont {Graham}, \citenamefont {Gu}, \citenamefont {Hughes},
  \citenamefont {Morr},\ and\ \citenamefont {Madhavan}}]{Madhavan2020}%
  \BibitemOpen
  \bibfield  {author} {\bibinfo {author} {\bibfnamefont {Z.}~\bibnamefont
  {Wang}}, \bibinfo {author} {\bibfnamefont {J.~O.}\ \bibnamefont {Rodriguez}},
  \bibinfo {author} {\bibfnamefont {L.}~\bibnamefont {Jiao}}, \bibinfo {author}
  {\bibfnamefont {S.}~\bibnamefont {Howard}}, \bibinfo {author} {\bibfnamefont
  {M.}~\bibnamefont {Graham}}, \bibinfo {author} {\bibfnamefont
  {G.}~\bibnamefont {Gu}}, \bibinfo {author} {\bibfnamefont {T.~L.}\
  \bibnamefont {Hughes}}, \bibinfo {author} {\bibfnamefont {D.~K.}\
  \bibnamefont {Morr}},\ and\ \bibinfo {author} {\bibfnamefont
  {V.}~\bibnamefont {Madhavan}},\ }\bibfield  {title} {\bibinfo {title}
  {Evidence for dispersing 1d majorana channels in an iron-based
  superconductor},\ }\href {https://doi.org/10.1126/science.aaw8419} {\bibfield
   {journal} {\bibinfo  {journal} {Science}\ }\textbf {\bibinfo {volume}
  {367}},\ \bibinfo {pages} {104} (\bibinfo {year} {2020})}\BibitemShut
  {NoStop}%
\bibitem [{\citenamefont {Zaki}\ \emph {et~al.}(2021)\citenamefont {Zaki},
  \citenamefont {Gu}, \citenamefont {Tsvelik}, \citenamefont {Wu},\ and\
  \citenamefont {Johnson}}]{Johnson}%
  \BibitemOpen
  \bibfield  {author} {\bibinfo {author} {\bibfnamefont {N.}~\bibnamefont
  {Zaki}}, \bibinfo {author} {\bibfnamefont {G.}~\bibnamefont {Gu}}, \bibinfo
  {author} {\bibfnamefont {A.}~\bibnamefont {Tsvelik}}, \bibinfo {author}
  {\bibfnamefont {C.}~\bibnamefont {Wu}},\ and\ \bibinfo {author}
  {\bibfnamefont {P.~D.}\ \bibnamefont {Johnson}},\ }\bibfield  {title}
  {\bibinfo {title} {Time-reversal symmetry breaking in the fe-chalcogenide
  superconductors},\ }\href {https://doi.org/10.1073/pnas.2007241118}
  {\bibfield  {journal} {\bibinfo  {journal} {Proceedings of the National
  Academy of Sciences}\ }\textbf {\bibinfo {volume} {118}},\ \bibinfo {pages}
  {e2007241118} (\bibinfo {year} {2021})}\BibitemShut {NoStop}%
\bibitem [{\citenamefont {Song}\ \emph {et~al.}(2022)\citenamefont {Song},
  \citenamefont {Zhang},\ and\ \citenamefont {Hao}}]{Song2022}%
  \BibitemOpen
  \bibfield  {author} {\bibinfo {author} {\bibfnamefont {R.}~\bibnamefont
  {Song}}, \bibinfo {author} {\bibfnamefont {P.}~\bibnamefont {Zhang}},\ and\
  \bibinfo {author} {\bibfnamefont {N.}~\bibnamefont {Hao}},\ }\bibfield
  {title} {\bibinfo {title} {Phase-manipulation-induced majorana mode and
  braiding realization in iron-based superconductor fe(te,se)},\ }\href
  {https://doi.org/10.1103/PhysRevLett.128.016402} {\bibfield  {journal}
  {\bibinfo  {journal} {Phys. Rev. Lett.}\ }\textbf {\bibinfo {volume} {128}},\
  \bibinfo {pages} {016402} (\bibinfo {year} {2022})}\BibitemShut {NoStop}%
\bibitem [{\citenamefont {Christensen}\ \emph {et~al.}(2019)\citenamefont
  {Christensen}, \citenamefont {Kang},\ and\ \citenamefont
  {Fernandes}}]{Rafael2019}%
  \BibitemOpen
  \bibfield  {author} {\bibinfo {author} {\bibfnamefont {M.~H.}\ \bibnamefont
  {Christensen}}, \bibinfo {author} {\bibfnamefont {J.}~\bibnamefont {Kang}},\
  and\ \bibinfo {author} {\bibfnamefont {R.~M.}\ \bibnamefont {Fernandes}},\
  }\bibfield  {title} {\bibinfo {title} {Intertwined spin-orbital coupled
  orders in the iron-based superconductors},\ }\href
  {https://doi.org/10.1103/PhysRevB.100.014512} {\bibfield  {journal} {\bibinfo
   {journal} {Phys. Rev. B}\ }\textbf {\bibinfo {volume} {100}},\ \bibinfo
  {pages} {014512} (\bibinfo {year} {2019})}\BibitemShut {NoStop}%
\bibitem [{\citenamefont {Mascot}\ \emph {et~al.}(2022)\citenamefont {Mascot},
  \citenamefont {Cocklin}, \citenamefont {Graham}, \citenamefont {Mashkoori},
  \citenamefont {Rachel},\ and\ \citenamefont {Morr}}]{Mascot2022}%
  \BibitemOpen
  \bibfield  {author} {\bibinfo {author} {\bibfnamefont {E.}~\bibnamefont
  {Mascot}}, \bibinfo {author} {\bibfnamefont {S.}~\bibnamefont {Cocklin}},
  \bibinfo {author} {\bibfnamefont {M.}~\bibnamefont {Graham}}, \bibinfo
  {author} {\bibfnamefont {M.}~\bibnamefont {Mashkoori}}, \bibinfo {author}
  {\bibfnamefont {S.}~\bibnamefont {Rachel}},\ and\ \bibinfo {author}
  {\bibfnamefont {D.~K.}\ \bibnamefont {Morr}},\ }\bibfield  {title} {\bibinfo
  {title} {Topological surface superconductivity in fese0. 45te0. 55},\ }\href
  {https://doi.org/10.1038/s42005-022-00943-x} {\bibfield  {journal} {\bibinfo
  {journal} {Communications Physics}\ }\textbf {\bibinfo {volume} {5}},\
  \bibinfo {pages} {188} (\bibinfo {year} {2022})}\BibitemShut {NoStop}%
\bibitem [{\citenamefont {Sarkar}\ \emph {et~al.}(2017)\citenamefont {Sarkar},
  \citenamefont {Van~Dyke}, \citenamefont {Sprau}, \citenamefont {Massee},
  \citenamefont {Welp}, \citenamefont {Kwok}, \citenamefont {Davis},\ and\
  \citenamefont {Morr}}]{Sarkar2017}%
  \BibitemOpen
  \bibfield  {author} {\bibinfo {author} {\bibfnamefont {S.}~\bibnamefont
  {Sarkar}}, \bibinfo {author} {\bibfnamefont {J.}~\bibnamefont {Van~Dyke}},
  \bibinfo {author} {\bibfnamefont {P.~O.}\ \bibnamefont {Sprau}}, \bibinfo
  {author} {\bibfnamefont {F.}~\bibnamefont {Massee}}, \bibinfo {author}
  {\bibfnamefont {U.}~\bibnamefont {Welp}}, \bibinfo {author} {\bibfnamefont
  {W.-K.}\ \bibnamefont {Kwok}}, \bibinfo {author} {\bibfnamefont {J.~C.~S.}\
  \bibnamefont {Davis}},\ and\ \bibinfo {author} {\bibfnamefont {D.~K.}\
  \bibnamefont {Morr}},\ }\bibfield  {title} {\bibinfo {title} {Orbital
  superconductivity, defects, and pinned nematic fluctuations in the doped iron
  chalcogenide ${\mathrm{fese}}_{0.45}{\mathrm{te}}_{0.55}$},\ }\href
  {https://doi.org/10.1103/PhysRevB.96.060504} {\bibfield  {journal} {\bibinfo
  {journal} {Phys. Rev. B}\ }\textbf {\bibinfo {volume} {96}},\ \bibinfo
  {pages} {060504} (\bibinfo {year} {2017})}\BibitemShut {NoStop}%
\bibitem [{\citenamefont {Fu}\ and\ \citenamefont {Kane}(2008)}]{Kane2008}%
  \BibitemOpen
  \bibfield  {author} {\bibinfo {author} {\bibfnamefont {L.}~\bibnamefont
  {Fu}}\ and\ \bibinfo {author} {\bibfnamefont {C.~L.}\ \bibnamefont {Kane}},\
  }\bibfield  {title} {\bibinfo {title} {Superconducting proximity effect and
  majorana fermions at the surface of a topological insulator},\ }\href
  {https://doi.org/10.1103/PhysRevLett.100.096407} {\bibfield  {journal}
  {\bibinfo  {journal} {Phys. Rev. Lett.}\ }\textbf {\bibinfo {volume} {100}},\
  \bibinfo {pages} {096407} (\bibinfo {year} {2008})}\BibitemShut {NoStop}%
\bibitem [{\citenamefont {Liu}\ \emph {et~al.}(2019)\citenamefont {Liu},
  \citenamefont {Liu}, \citenamefont {Zhang},\ and\ \citenamefont
  {Chiu}}]{Liu2019}%
  \BibitemOpen
  \bibfield  {author} {\bibinfo {author} {\bibfnamefont {C.-X.}\ \bibnamefont
  {Liu}}, \bibinfo {author} {\bibfnamefont {D.~E.}\ \bibnamefont {Liu}},
  \bibinfo {author} {\bibfnamefont {F.-C.}\ \bibnamefont {Zhang}},\ and\
  \bibinfo {author} {\bibfnamefont {C.-K.}\ \bibnamefont {Chiu}},\ }\bibfield
  {title} {\bibinfo {title} {Protocol for reading out majorana vortex qubits
  and testing non-abelian statistics},\ }\href
  {https://doi.org/10.1103/PhysRevApplied.12.054035} {\bibfield  {journal}
  {\bibinfo  {journal} {Phys. Rev. Appl.}\ }\textbf {\bibinfo {volume} {12}},\
  \bibinfo {pages} {054035} (\bibinfo {year} {2019})}\BibitemShut {NoStop}%
\bibitem [{\citenamefont {Fu}(2010)}]{Fu2010}%
  \BibitemOpen
  \bibfield  {author} {\bibinfo {author} {\bibfnamefont {L.}~\bibnamefont
  {Fu}},\ }\bibfield  {title} {\bibinfo {title} {Electron teleportation via
  majorana bound states in a mesoscopic superconductor},\ }\href
  {https://doi.org/10.1103/PhysRevLett.104.056402} {\bibfield  {journal}
  {\bibinfo  {journal} {Phys. Rev. Lett.}\ }\textbf {\bibinfo {volume} {104}},\
  \bibinfo {pages} {056402} (\bibinfo {year} {2010})}\BibitemShut {NoStop}%
\bibitem [{\citenamefont {Plugge}\ \emph {et~al.}(2016)\citenamefont {Plugge},
  \citenamefont {Zazunov}, \citenamefont {Eriksson}, \citenamefont {Tsvelik},\
  and\ \citenamefont {Egger}}]{Plugge2}%
  \BibitemOpen
  \bibfield  {author} {\bibinfo {author} {\bibfnamefont {S.}~\bibnamefont
  {Plugge}}, \bibinfo {author} {\bibfnamefont {A.}~\bibnamefont {Zazunov}},
  \bibinfo {author} {\bibfnamefont {E.}~\bibnamefont {Eriksson}}, \bibinfo
  {author} {\bibfnamefont {A.~M.}\ \bibnamefont {Tsvelik}},\ and\ \bibinfo
  {author} {\bibfnamefont {R.}~\bibnamefont {Egger}},\ }\bibfield  {title}
  {\bibinfo {title} {Kondo physics from quasiparticle poisoning in majorana
  devices},\ }\href {https://doi.org/10.1103/PhysRevB.93.104524} {\bibfield
  {journal} {\bibinfo  {journal} {Phys. Rev. B}\ }\textbf {\bibinfo {volume}
  {93}},\ \bibinfo {pages} {104524} (\bibinfo {year} {2016})}\BibitemShut
  {NoStop}%
\bibitem [{\citenamefont {Galpin}\ \emph {et~al.}(2014)\citenamefont {Galpin},
  \citenamefont {Mitchell}, \citenamefont {Temaismithi}, \citenamefont {Logan},
  \citenamefont {B\'eri},\ and\ \citenamefont {Cooper}}]{Galpin2014}%
  \BibitemOpen
  \bibfield  {author} {\bibinfo {author} {\bibfnamefont {M.~R.}\ \bibnamefont
  {Galpin}}, \bibinfo {author} {\bibfnamefont {A.~K.}\ \bibnamefont
  {Mitchell}}, \bibinfo {author} {\bibfnamefont {J.}~\bibnamefont
  {Temaismithi}}, \bibinfo {author} {\bibfnamefont {D.~E.}\ \bibnamefont
  {Logan}}, \bibinfo {author} {\bibfnamefont {B.}~\bibnamefont {B\'eri}},\ and\
  \bibinfo {author} {\bibfnamefont {N.~R.}\ \bibnamefont {Cooper}},\ }\bibfield
   {title} {\bibinfo {title} {Conductance fingerprint of majorana fermions in
  the topological kondo effect},\ }\href
  {https://doi.org/10.1103/PhysRevB.89.045143} {\bibfield  {journal} {\bibinfo
  {journal} {Phys. Rev. B}\ }\textbf {\bibinfo {volume} {89}},\ \bibinfo
  {pages} {045143} (\bibinfo {year} {2014})}\BibitemShut {NoStop}%
\bibitem [{\citenamefont {Zazunov}\ \emph {et~al.}(2014)\citenamefont
  {Zazunov}, \citenamefont {Altland},\ and\ \citenamefont
  {Egger}}]{zazunov2014transport}%
  \BibitemOpen
  \bibfield  {author} {\bibinfo {author} {\bibfnamefont {A.}~\bibnamefont
  {Zazunov}}, \bibinfo {author} {\bibfnamefont {A.}~\bibnamefont {Altland}},\
  and\ \bibinfo {author} {\bibfnamefont {R.}~\bibnamefont {Egger}},\ }\bibfield
   {title} {\bibinfo {title} {Transport properties of the coulomb--majorana
  junction},\ }\href@noop {} {\bibfield  {journal} {\bibinfo  {journal} {New
  Journal of Physics}\ }\textbf {\bibinfo {volume} {16}},\ \bibinfo {pages}
  {015010} (\bibinfo {year} {2014})}\BibitemShut {NoStop}%
\bibitem [{\citenamefont {Audouard}\ \emph {et~al.}(2015)\citenamefont
  {Audouard}, \citenamefont {Duc}, \citenamefont {Drigo}, \citenamefont
  {Toulemonde}, \citenamefont {Karlsson}, \citenamefont {Strobel},\ and\
  \citenamefont {Sulpice}}]{audouard2015quantum}%
  \BibitemOpen
  \bibfield  {author} {\bibinfo {author} {\bibfnamefont {A.}~\bibnamefont
  {Audouard}}, \bibinfo {author} {\bibfnamefont {F.}~\bibnamefont {Duc}},
  \bibinfo {author} {\bibfnamefont {L.}~\bibnamefont {Drigo}}, \bibinfo
  {author} {\bibfnamefont {P.}~\bibnamefont {Toulemonde}}, \bibinfo {author}
  {\bibfnamefont {S.}~\bibnamefont {Karlsson}}, \bibinfo {author}
  {\bibfnamefont {P.}~\bibnamefont {Strobel}},\ and\ \bibinfo {author}
  {\bibfnamefont {A.}~\bibnamefont {Sulpice}},\ }\bibfield  {title} {\bibinfo
  {title} {Quantum oscillations and upper critical magnetic field of the
  iron-based superconductor fese},\ }\href@noop {} {\bibfield  {journal}
  {\bibinfo  {journal} {Europhysics Letters}\ }\textbf {\bibinfo {volume}
  {109}},\ \bibinfo {pages} {27003} (\bibinfo {year} {2015})}\BibitemShut
  {NoStop}%
\bibitem [{\citenamefont {Singh}\ \emph {et~al.}(2015)\citenamefont {Singh},
  \citenamefont {White}, \citenamefont {Schmaus}, \citenamefont {Tsurkan},
  \citenamefont {Loidl}, \citenamefont {Deisenhofer},\ and\ \citenamefont
  {Wahl}}]{singh2015evidence}%
  \BibitemOpen
  \bibfield  {author} {\bibinfo {author} {\bibfnamefont {U.~R.}\ \bibnamefont
  {Singh}}, \bibinfo {author} {\bibfnamefont {S.~C.}\ \bibnamefont {White}},
  \bibinfo {author} {\bibfnamefont {S.}~\bibnamefont {Schmaus}}, \bibinfo
  {author} {\bibfnamefont {V.}~\bibnamefont {Tsurkan}}, \bibinfo {author}
  {\bibfnamefont {A.}~\bibnamefont {Loidl}}, \bibinfo {author} {\bibfnamefont
  {J.}~\bibnamefont {Deisenhofer}},\ and\ \bibinfo {author} {\bibfnamefont
  {P.}~\bibnamefont {Wahl}},\ }\bibfield  {title} {\bibinfo {title} {Evidence
  for orbital order and its relation to superconductivity in fese0. 4te0. 6},\
  }\href@noop {} {\bibfield  {journal} {\bibinfo  {journal} {Science Advances}\
  }\textbf {\bibinfo {volume} {1}},\ \bibinfo {pages} {e1500206} (\bibinfo
  {year} {2015})}\BibitemShut {NoStop}%
\bibitem [{\citenamefont {Eriksson}\ \emph {et~al.}(2014)\citenamefont
  {Eriksson}, \citenamefont {Nava}, \citenamefont {Mora},\ and\ \citenamefont
  {Egger}}]{ErikssonPRB2014}%
  \BibitemOpen
  \bibfield  {author} {\bibinfo {author} {\bibfnamefont {E.}~\bibnamefont
  {Eriksson}}, \bibinfo {author} {\bibfnamefont {A.}~\bibnamefont {Nava}},
  \bibinfo {author} {\bibfnamefont {C.}~\bibnamefont {Mora}},\ and\ \bibinfo
  {author} {\bibfnamefont {R.}~\bibnamefont {Egger}},\ }\bibfield  {title}
  {\bibinfo {title} {Tunneling spectroscopy of majorana-kondo devices},\ }\href
  {https://doi.org/10.1103/PhysRevB.90.245417} {\bibfield  {journal} {\bibinfo
  {journal} {Phys. Rev. B}\ }\textbf {\bibinfo {volume} {90}},\ \bibinfo
  {pages} {245417} (\bibinfo {year} {2014})}\BibitemShut {NoStop}%
\bibitem [{\citenamefont {Gabay}\ \emph {et~al.}(2022)\citenamefont {Gabay},
  \citenamefont {Han}, \citenamefont {Lopes}, \citenamefont {Affleck},\ and\
  \citenamefont {Sela}}]{Sela2022}%
  \BibitemOpen
  \bibfield  {author} {\bibinfo {author} {\bibfnamefont {D.}~\bibnamefont
  {Gabay}}, \bibinfo {author} {\bibfnamefont {C.}~\bibnamefont {Han}}, \bibinfo
  {author} {\bibfnamefont {P.~L.~S.}\ \bibnamefont {Lopes}}, \bibinfo {author}
  {\bibfnamefont {I.}~\bibnamefont {Affleck}},\ and\ \bibinfo {author}
  {\bibfnamefont {E.}~\bibnamefont {Sela}},\ }\bibfield  {title} {\bibinfo
  {title} {Multi-impurity chiral kondo model: Correlation functions and anyon
  fusion rules},\ }\href {https://doi.org/10.1103/PhysRevB.105.035151}
  {\bibfield  {journal} {\bibinfo  {journal} {Phys. Rev. B}\ }\textbf {\bibinfo
  {volume} {105}},\ \bibinfo {pages} {035151} (\bibinfo {year}
  {2022})}\BibitemShut {NoStop}%
\bibitem [{\citenamefont {Lotem}\ \emph {et~al.}(2022)\citenamefont {Lotem},
  \citenamefont {Sela},\ and\ \citenamefont {Goldstein}}]{Lotem2022}%
  \BibitemOpen
  \bibfield  {author} {\bibinfo {author} {\bibfnamefont {M.}~\bibnamefont
  {Lotem}}, \bibinfo {author} {\bibfnamefont {E.}~\bibnamefont {Sela}},\ and\
  \bibinfo {author} {\bibfnamefont {M.}~\bibnamefont {Goldstein}},\ }\bibfield
  {title} {\bibinfo {title} {Manipulating non-abelian anyons in a chiral
  multichannel kondo model},\ }\href
  {https://doi.org/10.1103/PhysRevLett.129.227703} {\bibfield  {journal}
  {\bibinfo  {journal} {Phys. Rev. Lett.}\ }\textbf {\bibinfo {volume} {129}},\
  \bibinfo {pages} {227703} (\bibinfo {year} {2022})}\BibitemShut {NoStop}%
\bibitem [{\citenamefont {Ying}\ \emph {et~al.}(2022)\citenamefont {Ying},
  \citenamefont {Zhang}, \citenamefont {Chen}, \citenamefont {Jia},
  \citenamefont {Fei}, \citenamefont {Zhang}, \citenamefont {Zhang},
  \citenamefont {Wang},\ and\ \citenamefont {Song}}]{MnBiTe}%
  \BibitemOpen
  \bibfield  {author} {\bibinfo {author} {\bibfnamefont {Z.}~\bibnamefont
  {Ying}}, \bibinfo {author} {\bibfnamefont {S.}~\bibnamefont {Zhang}},
  \bibinfo {author} {\bibfnamefont {B.}~\bibnamefont {Chen}}, \bibinfo {author}
  {\bibfnamefont {B.}~\bibnamefont {Jia}}, \bibinfo {author} {\bibfnamefont
  {F.}~\bibnamefont {Fei}}, \bibinfo {author} {\bibfnamefont {M.}~\bibnamefont
  {Zhang}}, \bibinfo {author} {\bibfnamefont {H.}~\bibnamefont {Zhang}},
  \bibinfo {author} {\bibfnamefont {X.}~\bibnamefont {Wang}},\ and\ \bibinfo
  {author} {\bibfnamefont {F.}~\bibnamefont {Song}},\ }\bibfield  {title}
  {\bibinfo {title} {Experimental evidence for dissipationless transport of the
  chiral edge state of the high-field chern insulator in
  ${\mathrm{mnbi}}_{2}{\mathrm{te}}_{4}$ nanodevices},\ }\href
  {https://doi.org/10.1103/PhysRevB.105.085412} {\bibfield  {journal} {\bibinfo
   {journal} {Phys. Rev. B}\ }\textbf {\bibinfo {volume} {105}},\ \bibinfo
  {pages} {085412} (\bibinfo {year} {2022})}\BibitemShut {NoStop}%
\end{thebibliography}%

\end{document}